\documentclass[10pt,numbers]{sigplanconf}

\usepackage{bigfoot} 

\usepackage{amsmath,bm,amssymb,stmaryrd}
\usepackage{amsthm}
\usepackage{ifthen,plstx,mathpartir, enumerate}
\usepackage[usenames,dvipsnames]{xcolor}
\usepackage{titlesec}
\usepackage{tcolorbox}

\usepackage{tikz}

\usepackage{natbib}

\usepackage[firstpage]{draftwatermark}
\SetWatermarkText{\hspace*{7in}\raisebox{7.75in}{\includegraphics[scale=0.3]{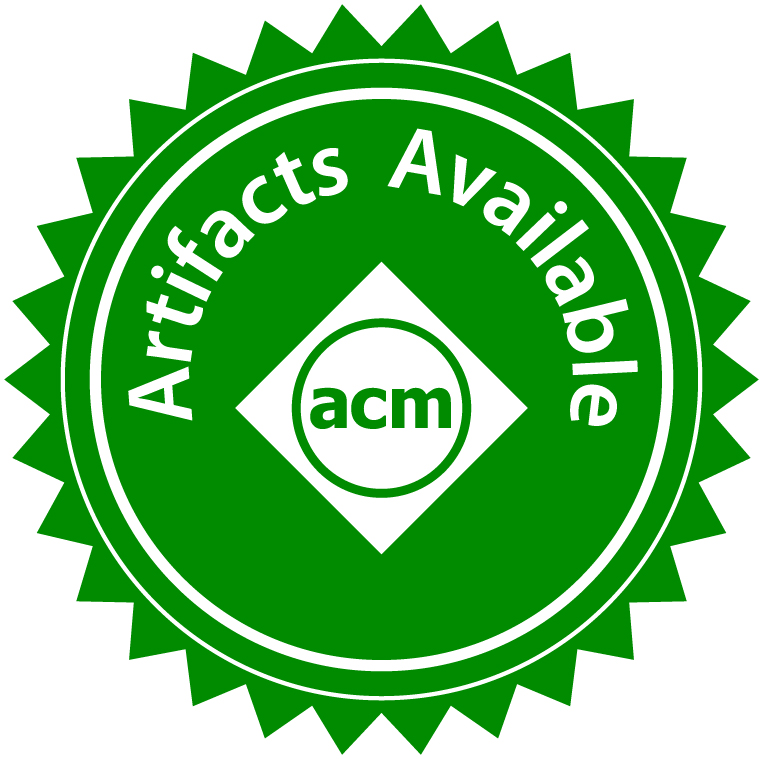}\includegraphics[scale=0.3]{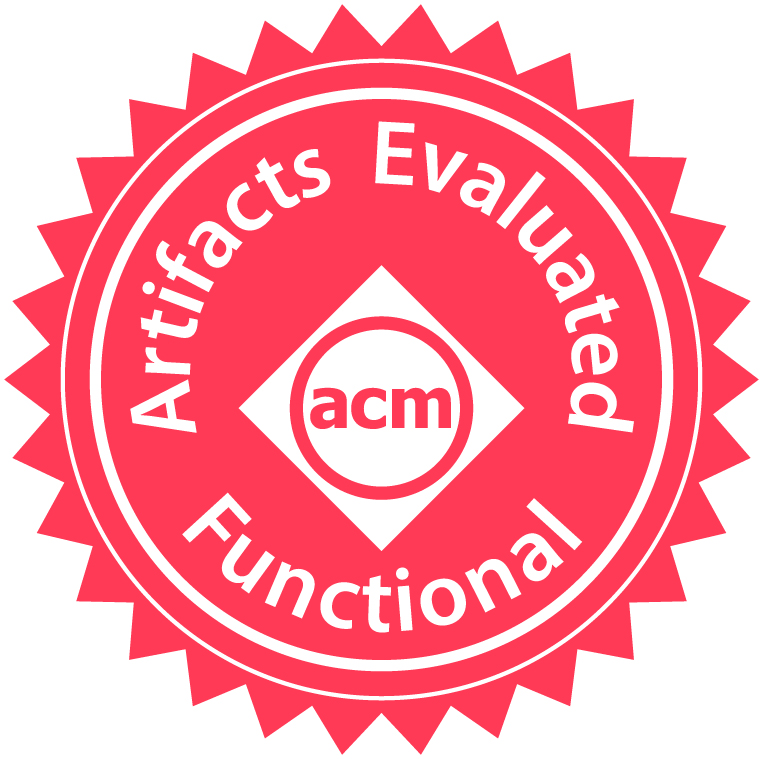}}}
\SetWatermarkAngle{0}

\newcommand{\omitthis}[1]{}

\newcommand{\etal}{\textit{et al.}}

\renewcommand{\mit}[1]{\mathit{#1}}
\newcommand{\mrm}[1]{\mathrm{#1}}

\newenvironment{nop}{}{}

\newenvironment{stackAux}[2]{%
\setlength{\arraycolsep}{0pt}
\begin{array}[#1]{#2}}{
\end{array}}

\newenvironment{stackTL}{
\begin{stackAux}{t}{l}}{\end{stackAux}}

\newtheoremstyle{athm}{\topsep}{\topsep}%
{}
     {}
     {\bfseries}
     {}
     {\newline}
     {\thmname{#1}\thmnumber{ #2}\thmnote{~\,(#3)}}

\theoremstyle{athm}
\newtheorem{theorem}{Theorem}[section]

\renewcommand{\arraystretch}{1.3}

\newcommand{\SRC}{F}
\newcommand{\CLO}{C}
\newcommand{\ALC}{A}
\newcommand{\TAL}{T}
\newcommand{\ANF}{Z}

\definecolor{orange}{cmyk}{0,0.5,1,.3}

\newcommand{\langcolor}[2]
   {\ifthenelse{\equal{#1}{\SRC}}{{\color{RoyalBlue} #2}}
   {\ifthenelse{\equal{#1}{\ANF}}{{\color{Orange} #2}}
   {\ifthenelse{\equal{#1}{\CLO}}{{\color{red} #2}}
   {\ifthenelse{\equal{#1}{\ALC}}{{\color{blue} #2}}
   {\ifthenelse{\equal{#1}{\TAL}}{{\color{OrangeRed} #2}}
   {{\color{black} #2}}}}}}}

\newcommand{\langfont}[2]
   {\mbox{\unboldmath{\ensuremath
   {\ifthenelse{\equal{#1}{\SRC}}{{\bm{\mathsf{#2}}}}
   {\ifthenelse{\equal{#1}{\CLO}}{{\bm{\mathrm{#2}}}}
   {\ifthenelse{\equal{#1}{\ALC}}{{\bm{\mathsf{#2}}}}
   {\ifthenelse{\equal{#1}{\TAL}}{{\bm{\mathrm{#2}}}}
   {\ifthenelse{\equal{#1}{\ANF}}{{\mathit{#2}}}
   {#2}}}}}}}}}

\newcommand{\inlang}[2]{\ensuremath{\langcolor{#1}{\langfont{#1}{{#2}}}}}

\newcommand{\src}{\inlang\SRC}

\newcommand{\tal}{\inlang\TAL}

\newcommand{\bound}[5]
   {\ifthenelse{\equal{#1}{<}}
               {{^{\mbox{\scriptsize\inlang{#2}{#4}}\!}}
                \inlang0{\mathcal{#2#3}}\, \inlang{#3}{#5}}
   {\ifthenelse{\equal{#1}{>}}
               {\inlang0{\mathcal{#2#3}}
                ^{\mbox{\scriptsize\inlang{#3}{#4}}}\, \inlang{#3}{#5}}
   {\inlang0{\texttt{bad boundary syntax}}}}}

\newcommand{\trans}[3]
           {\inlang{#1}{#3}^{\mbox{\scriptsize\inlang{#2}{\mathcal{#2}}}}}

\newcommand{\tytrans}[3]
           {\inlang{#1}{#3}
            ^{\mbox{\scriptsize\inlang{#2}{\mathcal{#2}}}}}

\newcommand{\transval}[5]
   {\ifthenelse{\equal{#1}{<}}
               {{^{\mbox{\scriptsize\inlang{#2}{#4}}\!}}
                \inlang0{\textbf{\small #2#3}(\inlang{#3}{#5})}}
   {\ifthenelse{\equal{#1}{>}}
               {\inlang0{\textbf{\small #2#3}
                ^{\mbox{\scriptsize\inlang{#3}{#4}}}(\inlang{#3}{#5})}}
   {\inlang0{\texttt{bad boundary syntax}}}}}

\newcommand{\transhval}[5]
   {\ifthenelse{\equal{#1}{<}}
               {{^{\mbox{\scriptsize\inlang{#2}{#4}}\!}}
                \inlang0{\textbf{\small #2#3}_h(\inlang{#3}{#5})}}
   {\ifthenelse{\equal{#1}{>}}
               {\inlang0{\textbf{\small #2#3}_h
                ^{\mbox{\scriptsize\inlang{#3}{#4}}}(\inlang{#3}{#5})}}
   {\inlang0{\texttt{bad boundary syntax}}}}}

\newcommand{\withmem}[1]{\inlang0{{}, #1}}

\renewcommand{\int}{int}
\newcommand{\unitt}{unit} 
\newcommand{\arrow}[2]{(#1) \!\to #2}
\newcommand{\refto}[1]{ref\, #1}
\newcommand{\boxof}[1]{box\, #1}
\newcommand{\reftag}{{ref}}
\newcommand{\boxtag}{{box}}

\newcommand{\tuplet}[1]{\langle #1 \rangle} 
\newcommand{\pref}{\phi}

\newcommand{\emptystackt}{\bullet}
\newcommand{\codet}[4]{\forall [#1].\{#2; #3 \}^{#4}}

\newcommand{\stopt}[4]{\codet{}{#1 \: #2}{#3}{#4}}

\newcommand{\retmark}[2]{\tal{end\{{#1; #2}\}}}
\newcommand{\externmark}{\tal{out}}

\renewcommand{\:}{\colon\!}
\renewcommand{\vec}{\overline}
\newcommand{\unit}{()}
\newcommand{\sfunc}[2]{\lambda(#1).#2}
\newcommand{\func}[3][]{\lambda^{\tal{#1}}(#2).#3}
\newcommand{\sapp}[2]{{#1}\, {#2}}
\newcommand{\app}[2]{{#1}\, {#2}}
\newcommand{\pack}[3]{pack \langle #1,\! #2 \rangle\, as\, #3}

\newcommand{\fold}[2]{fold_{#1}\, #2}
\newcommand{\unfold}[1]{unfold\, #1}

\newcommand{\proj}[2]{\pi_{#1}(#2)}
\newcommand{\test}[3]{if0\, {#1}\ {#2}\ {#3}}

\newcommand{\hole}{[\cdot]}

\newcommand{\tuple}[1]{\langle #1 \rangle}

\newcommand{\reg}[1]{\mathtt{r#1}}

\newcommand{\code}[5]{code [#1]\{#2; #3\}^{#4}.#5}
\newcommand{\codesplit}[5]{code [#1]\{\begin{stackTL}#2;\\ #3\}^{#4}.\end{stackTL}#5}

\newcommand{\aop}{aop}

\newcommand{\mult}{\mathtt{mult}}
\newcommand{\jmp}[1]{\mathtt{jmp}\, #1}

\newcommand{\binop}[4]{#1\ #2, #3, #4}
\newcommand{\bnz}[2]{\mathtt{bnz}\, #1, #2}
\newcommand{\ld}[3]{\mathtt{ld}\, #1, #2 [#3]}
\newcommand{\st}[3]{\mathtt{st}\, #1 [#2], #3}

\newcommand{\refallocinstr}[2]{\mathtt{ralloc}\, #1, #2}
\newcommand{\boxallocinstr}[2]{\mathtt{balloc}\, #1, #2}
\newcommand{\mv}[2]{\mathtt{mv}\, #1, #2}
\newcommand{\unpackinstr}[3]{\mathtt{unpack}\, \langle #1, #2 \rangle\ #3}
\newcommand{\unfoldinstr}[2]{\mathtt{unfold}\, #1, #2}
\newcommand{\salloc}[1]{\mathtt{salloc}\, #1}
\newcommand{\sfree}[1]{\mathtt{sfree}\, #1}
\newcommand{\sld}[2]{\mathtt{sld}\, #1, #2}
\newcommand{\sst}[2]{\mathtt{sst}\, #1, #2}
\newcommand{\emptystack}{nil}
\newcommand{\cons}[2]{#1 :: #2}

\newcommand{\import}[3]{\mathtt{import}\ #1, {}^{#2}{#3}}
\newcommand{\protec}[2]{\mathtt{protect}\ {\tal{#1}}, {\tal{#2}}}

\newcommand{\callinstr}[3]{\mathtt{call}\, #1\, \{#2, #3\}}
\newcommand{\call}{\callinstr}
\newcommand{\retinstr}[2]{\mathtt{ret}\, #1\, \{#2\}}
\newcommand{\haltinstr}[3]{\mathtt{halt}\, #1, #2\, \{#3\}}

\newcommand{\sfuncz}[4]{\lambda^{\tal{#1}}_{\tal{#2}}(#3).#4}
\newcommand{\sarrow}[4]{(#1) \overset{\tal{#2}; \tal{#3}}{\longrightarrow} #4}

\newcommand{\wft}[3][]{\ifthenelse{\equal{#1}{}}
                         {\ensuremath{#2 \vdash #3}}
                         {\ensuremath{\inlang{#1}{#2} \vdash \inlang{#1}{#3}}}}
\newcommand{\wftagt}[4][]{\ifthenelse{\equal{#1}{}}
                         {\ensuremath{#2 \vdash^{#3} #4}}
                         {\ensuremath{\inlang{#1}{#2} \vdash^{\inlang{#1}{#3}}
                                      \inlang{#1}{#4}}}}
\newcommand{\wfenv}[3][]{\ifthenelse{\equal{#1}{}}
                         {\ensuremath{#2 \vdash #3}}
                         {\ensuremath{\inlang{#1}{#2} \vdash
                                      \inlang{#1}{#3}}}}
\newcommand{\tmemfrag}[4][]{\ifthenelse{\equal{#1}{}}
                         {\ensuremath{#2 \vdash #3 \: #4}}
                         {\ensuremath{\inlang{#1}{#2} \vdash
                                      \inlang{#1}{#3} \: \inlang{#1}{#4}}}}

\newcommand{\wfm}[3][]{\ifthenelse{\equal{#1}{}}
                         {\ensuremath{\vdash #2 \: #3}}
                         {\ensuremath{\vdash
                                      \inlang{#1}{#2} \: \inlang{#1}{#3}}}}

\newcommand{\wfret}[6][]{\ifthenelse{\equal{#1}{}}
                         {\ensuremath{#2[#3]; #4; #5 \vdash #6}}
                         {\ensuremath{\inlang{#1}{#2}\bm{[}\inlang{#1}{#3}\bm{]};
                                      \inlang{#1}{#4}; \inlang{#1}{#5}
                                      \vdash \inlang{#1}{#6}}}}

\newcommand{\tbasic}[4][]{\ifthenelse{\equal{#1}{}}
                         {\ensuremath{#2 \vdash #3 \: #4}}
                         {\ensuremath{\inlang{#1}{#2} \vdash
                                      \inlang{#1}{#3} \: \inlang{#1}{#4}}}}
\renewcommand{\t}[6][]{\ifthenelse{\equal{#1}{}}
                         {\ensuremath{#2; #3; #4 \vdash #5 \: #6}}
                         {\ensuremath{\inlang{#1}{#2}; \inlang{#1}{#3};
                                      \inlang{#1}{#4} \vdash
                                      \inlang{#1}{#5} \: \inlang{#1}{#6}}}}
\newcommand{\thval}[5][]{\ifthenelse{\equal{#1}{}}
                         {\ensuremath{#2 \vdash #3 \:^{#4} #5}}
                         {\ensuremath{\inlang{#1}{#2} \vdash
                                      \inlang{#1}{#3} \:^{\inlang{#1}{#4}} \inlang{#1}{#5}}}}
\newcommand{\twval}[5][]{\ifthenelse{\equal{#1}{}}
                         {\ensuremath{#2; #3 \vdash #4 \: #5}}
                         {\ensuremath{\inlang{#1}{#2}; \inlang{#1}{#3} \vdash
                                      \inlang{#1}{#4} \: \inlang{#1}{#5}}}}
\newcommand{\tval}[5][]{\ifthenelse{\equal{#1}{}}
                         {\ensuremath{#2; #3 \vdash #4 \: #5}}
                         {\ensuremath{\inlang{#1}{#2}; \inlang{#1}{#3} \vdash
                                      \inlang{#1}{#4} \: \inlang{#1}{#5}}}}
\newcommand{\tinstr}[8][]{\ifthenelse{\equal{#1}{}}
                         {\ensuremath{#2; #3; \tal{#4}; \tal{#5};
                                      \tal{#6} \vdash \tal{#7}
                                      \Rightarrow {#8}}}
                         {\ensuremath{\inlang{#1}{#2}; \inlang{#1}{#3};
                                      \inlang{#1}{#4}; \inlang{#1}{#5};
                                      \inlang{#1}{#6} \vdash \inlang{#1}{#7}
                                      \Rightarrow {#8}}}}
\newcommand{\tinstrbreak}[8][]{\ifthenelse{\equal{#1}{}}
                         {\ensuremath{#2; #3; \tal{#4}; \tal{#5};
                                      \tal{#6} \begin{stackTL}{} \vdash \tal{#7} \\
                                      \quad \Rightarrow {#8} \end{stackTL}}}
                         {\ensuremath{\inlang{#1}{#2}; \inlang{#1}{#3};
                                      \inlang{#1}{#4}; \inlang{#1}{#5};
                                      \inlang{#1}{#6} \vdash \inlang{#1}{#7}
                                      \Rightarrow {#8}}}}
\newcommand{\instrpost}[5][\TAL]{\ifthenelse{\equal{#1}{}}
                         {\ensuremath{#2; \tal{#3}; \tal{#4}; \tal{#5}}}
                         {\ensuremath{\inlang{#1}{#2}; \inlang{#1}{#3};
                                      \inlang{#1}{#4}; \inlang{#1}{#5}}}}
\newcommand{\tinseq}[7][]{\ifthenelse{\equal{#1}{}}
                         {\ensuremath{#2; #3; \tal{#4}; \tal{#5};
                                      \tal{#6} \vdash \tal{#7}}}
                         {\ensuremath{\inlang{#1}{#2}; \inlang{#1}{#3};
                                      \inlang{#1}{#4}; \inlang{#1}{#5};
                                      \inlang{#1}{#6} \vdash \inlang{#1}{#7}}}}
\newcommand{\subt}[4][]{\ifthenelse{\equal{#1}{}}
                         {\ensuremath{#2 \vdash #3 \le #4}}
                         {\ensuremath{\inlang{#1}{#2} \vdash
                                      \inlang{#1}{#3} \le \inlang{#1}{#4}}}}
\newcommand{\texpr}[9][]{\ifthenelse{\equal{#1}{}}
                         {\ensuremath{#2; #3; #4; #5; #6 \vdash #7 \: #8; #9}}
                         {\ensuremath{\inlang{#1}{#2}; \inlang{#1}{#3};
                                      \inlang{#1}{#4}; \inlang{#1}{#5};
                                      \inlang{#1}{#6} \vdash \inlang{#1}{#7} \:
                                      \inlang{#1}{#8}; \inlang{#1}{#9}}}}
\newcommand{\tfull}[8]{\ensuremath{#1;#2;\tal{#3};\tal{#4};\tal{#5} \vdash
                                        #6 \: #7; \tal{#8}}}

\newcommand{\red}[1][]{\inlang0{{}\longmapsto^{#1}{}}}

\DeclareMathOperator{\retreg}{ret-reg}

\newcommand{\extends}{\sqsupseteq}

\newcommand{\worldext}{\extends}
\newcommand{\pubextends}{\extends_{\mrm{pub}}}
\newcommand{\worldextpub}{\pubextends}
\newcommand{\later}{\rhd}

\newcommand{\obs}{\mathcal{O}}

\newcommand{\lraprx}{\approx}

\newcommand{\ctxeqv}{\approx^{\mit{ctx}}}

\DeclareMathOperator{\currS}{curr-S}
\DeclareMathOperator{\currR}{curr-R}

\newcommand{\interp}[2]{\mathcal{#1} \llbracket #2 \rrbracket}

\newcommand{\Erel}[3]{\interp E {\tal{#1} \vdash #2; \tal{#3}}}

\newcommand{\Krel}[3]{\interp K {\tal{#1} \vdash #2; \tal{#3}}}

\DeclareMathOperator{\typeof}{ret-type} 
\DeclareMathOperator{\retaddrtype}{ret-addr-type}
\DeclareMathOperator{\retaddr}{ret-addr}
\newcommand{\incq}[2]{\inlang0{\text{inc}(\tal{#1},\tal{#2})}}

\newcommand{\config}[2]{\langle#1 \mid #2\rangle}

\newcommand{\defeq}{\stackrel{\mathrm{def}}{=}}

\newcommand{\phiv}[1]
           {\ifthenelse{\equal{#1}{0}}{{\varphi_v}}{{\varphi_v^{#1}}}}
\newcommand{\phivhat}[1]
           {\ifthenelse{\equal{#1}{0}}{{\hat\varphi_v}}{{\hat\varphi_v^{#1}}}}

\newcommand{\tlang}{\tal{\mathbb{T}}}
\newcommand{\flang}{\src{\mathbb{F}}}
\newcommand{\ftlang}{\src{\mathbb{F}}\tal{\mathbb{T}}}

\newcommand{\alang}{\src{\mathbb{A}}}
\newcommand{\atlang}{\src{\mathbb{A}}\tal{\mathbb{T}}}
\newcommand{\sfactor}{1.1}

\newcommand{\cns}[2]{#1 \! :: \! #2}
\newcommand{\map}{\! \mapsto\!}
\newcommand{\calledge}[3]{\footnotesize $\tal{\mathtt{#1}}$ \\[-2pt] \scriptsize$\mathrm{\color{Gray}#2}$ \\[-4pt] \scriptsize$\mathrm{\color{Gray}#3}$}
\newcommand{\fcall}[1]{\scriptsize$\src{#1}$}

\tikzstyle{vertex}=[circle,fill=black,minimum size=5pt,inner sep=0pt]
\tikzstyle{blank}=[circle,minimum size=0pt,inner sep=0pt]
\tikzstyle{arrow} = [thick,->,>=stealth] 
\tikzstyle{memory} = [align=left, midway, sharp corners, thin, dashed, draw=black, fill=White, minimum width=0]

\tikzstyle{ftalblock} = [rectangle, ultra thick, rounded corners, minimum width=1cm, minimum height=0.5cm,text centered, draw=RoyalBlue!65, fill=OrangeRed!10]

\tikzstyle{talfblock} = [rectangle, ultra thick, rounded corners, minimum width=1cm, minimum height=0.5cm,text centered, draw=OrangeRed!65, fill=RoyalBlue!10]

\tikzstyle{talblock} = [rectangle, ultra thick, rounded corners, minimum width=1cm, minimum height=0.5cm,text centered, draw=OrangeRed!65, fill=OrangeRed!10]

\tikzstyle{fblock} = [rectangle, rounded corners, ultra thick, minimum width=1cm, minimum height=0.5cm,text centered, draw=RoyalBlue!65, fill=RoyalBlue!10]

\tikzstyle{genblock} = [rectangle, ultra thick, rounded corners, minimum width=1cm, minimum height=0.5cm,text centered, draw=Gray!75, fill=Gray!40]

\newtcbox{\graybox}{nobeforeafter,colframe=Gray!75,colback=Gray!40,boxrule=2pt,arc=2pt,
  boxsep=0pt,left=1pt,right=1pt,top=1pt,bottom=1pt,tcbox raise base}
\newtcbox{\tlangbox}{nobeforeafter,colframe=OrangeRed!75,colback=OrangeRed!10,boxrule=2pt,arc=2pt,
  boxsep=0pt,left=1pt,right=1pt,top=1pt,bottom=1pt,tcbox raise base}
\newtcbox{\flangbox}{nobeforeafter,colframe=RoyalBlue!75,colback=RoyalBlue!10,boxrule=2pt,arc=2pt,
  boxsep=0pt,left=1pt,right=1pt,top=1pt,bottom=1pt,tcbox raise base}

\tikzstyle{thalt}=[rectangle,draw=OrangeRed,fill=Purple,minimum size=4pt,inner sep=1pt]
\tikzstyle{fhalt}=[rectangle,draw=RoyalBlue,fill=RoyalBlue,minimum size=4pt,inner sep=1pt]

\usepackage{balance} 
\begin{document}
\toappear{}

\setlength{\pdfpageheight}{\paperheight}
\setlength{\pdfpagewidth}{\paperwidth}

\conferenceinfo{CONF 'yy}{Month d--d, 20yy, City, ST, Country}
\copyrightyear{20yy}
\copyrightdata{978-1-nnnn-nnnn-n/yy/mm}
\copyrightdoi{nnnnnnn.nnnnnnn}

\publicationrights{licensed}     

\titlebanner{PLDI'17 Draft}        
\preprintfooter{FunTAL: Reasonably mixing a functional language with assembly}   

\title{\protect{\src{\mathbb{F}}}un\protect{\tal{\mathbb{T}}}AL: Reasonably Mixing a Functional Language\\ with Assembly\thanks{We use $\src{blue}$ sans-serif to typeset our functional
  language $\flang$ and $\tal{red}$ roman to typeset our typed assembly 
  language $\tlang$. This paper will be much easier to follow if 
  read/printed in color.}}

 \authorinfo{Daniel Patterson}
            {Northeastern University, USA}
            {dbp@ccs.neu.edu}
\authorinfo{Jamie Perconti}
           {Northeastern University, USA}  
           {jamieperconti@gmail.com}
 \authorinfo{Christos Dimoulas}
            {Harvard University, USA}
            {chrdimo@seas.harvard.edu}
 \authorinfo{Amal Ahmed}
            {Northeastern University, USA}
            {amal@ccs.neu.edu}

\maketitle

\begin{abstract}
  We present FunTAL, the first multi-language system to formalize safe
interoperability between a high-level functional language and
low-level assembly code while supporting compositional reasoning about the
mix.  A central challenge in developing such a multi-language is
bridging the gap between assembly, which is staged into jumps to
continuations, 
and high-level code, where subterms return a result.
We present a \emph{compositional} stack-based typed assembly language
that supports \emph{components}, comprised of one or more basic blocks,
that may be embedded in high-level contexts.  We also present a
logical relation for FunTAL that supports reasoning about equivalence
of high-level components and their assembly replacements,
mixed-language programs with callbacks between languages, and assembly
components comprised of different numbers of basic blocks.


 
\end{abstract}

\category{F.3.1}{Logics and Meanings of Programs}{Specifying and Verifying and Reasoning about Programs}
\category{D.3.1}{Programming Languages}{Formal Definitions and Theory---Semantics}

\keywords
multi-language semantics, typed assembly language, inline assembly,
contextual equivalence, logical relations

\section{Introduction}
\label{sec:intro}

Developers frequently integrate code written in lower-level
languages into their high-level-language programs.  For instance,
OCaml and Haskell developers may leverage the FFI to make use of
libraries implemented in C, while Rust developers may include inline
assembly directly.  In each of these cases, developers resort to the
lower-level language so they can use features unavailable in the
high-level language to gain access to hardware or fine-tune performance.

However, the benefits of mixed-language programs come at a price. To reason
about the behavior of a high-level component, developers need to think not only
about the semantics of the high-level language, but also about the way their
high-level code was compiled and all interactions with low-level
code. Since low-level code usually comes without safety guarantees, invalid
instructions could crash the program. More insidiously, low-level code can
potentially alter control flow, mutate values that should be inaccessible, or
introduce security vulnerabilities that would not be possible in the high-level
language. Unfortunately, there are no mixed-language systems that enable
non-expert programmers to reason about interactions with lower-level
code---i.e., systems that guarantee safe interoperability and provide rules for
compositional reasoning in a mixed-language setting.

Even if developers don't directly write inline assembly,
mixed-language programs are a reality that compiler writers and
compiler-verification efforts must contend with.  For  
instance, mixed programs show up in modern just-in-time (JIT)  
compilers, where the high-level language is initially interpreted
until the runtime can identify portions to statically compile, at
which point those portions of the code are replaced with equivalent
assembly. These assembly components will include hooks to move back
into the interpreted runtime, corresponding closely to the semantics
of a mixed-language program.  Verifying correctness of such JITs
requires proving that the high-level fragment and its compiled replacement
are \emph{contextually equivalent} in the mixed language. Contextual
equivalence guarantees that in any whole program replacing the
high-level fragment with the compiled version will not change the
behavior of the program.   

In the case of traditional compilers, compiled components are
frequently linked with target code compiled from a different source
language, or with low-level routines that form part of the runtime
system.  
Perconti and Ahmed~\cite{perconti14:fca,ahmed15:snapl} argue that 
correctness theorems for verified compilers that account for such
linking must include mixed-language reasoning.
Specifically, they set up multi-languages that specify the rules of
source-target interoperability and then express compiler correctness as
multi-language equivalence between a source component $e_S$ and its
compiled version $e_T$.  Hence, the theorem ensures that $e_T$
linked with some arbitrary target code $e_T'$ will behave the same as
$e_S$ interoperating with $e_T'$. \looseness=-1

All of the above scenarios call for the design of a multi-language that
specifies interoperability between a high-level language and assembly,
along with proof methods for reasoning about equivalence of components
in this setting. Note that Perconti and
Ahmed~\cite{perconti14:fca} left the design of a multi-language that 
embeds assembly as future work. Since they did not show how to verify 
a code generation pass to assembly, they didn't need to define
interoperability between a high-level, expression-based language and a
low-level language with direct jumps. 

In this paper, we present FunTAL, a multi-language system that allows 
assembly to be embedded in a typed functional language and vice versa.
A key difficulty is ensuring that the embedded assembly has local and
well-controlled effects.  This is challenging because assembly is
inherently \emph{non-compositional}---control can change to an
arbitrary point with direct jumps and code can access arbitrary values
far up on the call-stack.  To allow a compositional functional
language to safely interoperate with assembly, such behavior must be
constrained, which we do using types at the assembly level.  Moreover,
we need to identify the right notion of \emph{component} in assembly:
intuitively, an assembly component may be comprised of multiple
basic blocks and we should be able to show equivalence between terms of 
the functional language (i.e., high-level components) and multi-block
assembly components.  But 
how do we identify which blocks should be grouped together into a
component without imposing so much high-level structure on assembly
that it ceases to be low level? Even once we identify such
groupings, we must still contend with the control-flow gap between a
direct-style functional language in which terms return results and
assembly code that is staged into jumps to continuations. Finally, we
must find a way to embed functional code in assembly so we can support
callbacks from assembly to the functional language.

\paragraph{Contributions}  We make the following contributions: 
\begin{itemize}
\vspace{-1ex}
\item 
We design a compositional typed assembly language (TAL) called
$\tlang$, building on the stack-based typed assembly language of
Morrisett~\etal~\cite{morrisett02} (henceforth, STAL).  The central
novelty of our TAL $\tlang$ are extensions to an STAL-like type system
that help us reason about multi-block components and bridge the gap between 
direct-style high-level components and continuation-based assembly
components (\S\ref{sec:tlang}). 
\item We present a multi-language $\ftlang$ in the style of
  Matthews-Findler~\cite{matthews07} that supports 
  interoperability between a simply typed functional language $\flang$
  with recursive types and our TAL $\tlang$
  (\S\ref{sec:ftlang}). 
\item We develop a novel step-indexed Kripke logical relation for
  reasoning about equivalence of $\ftlang$ components (\S\ref{sec:lr}).  It builds on
  prior logical relations for mutable
  state~\cite{ahmed09:sdri,dreyer12,perconti14:fca}, but is the first 
  to support reasoning about equivalence of programs that mix assembly
  with lambdas (including callbacks between them), and of assembly
  components comprised of different numbers of basic blocks.  The
  central novelty lies in the mechanics of  
  accommodating assembly and equivalence of multi-block components.  
\end{itemize}

\vspace{-0.8ex}

The technical appendix~\cite{patterson17:funtal-tr} includes complete
language semantics, definitions, and proofs, some of which are elided in
this paper. Our artifact provides an in-browser type checker and
machine stepper for the multi-language to aid understanding and 
experimentation with $\ftlang$ programs. The artifact, available at
\verb|https://dbp.io/artifacts/funtal|, includes runnable versions of
all examples in the paper.


\section{Main Ingredients of the Mix}
\label{sec:main-ideas}

We design a \emph{compositional} TAL $\tlang$ that draws largely from
Morrisett~\etal's STAL~\cite{morrisett02}, which has a single explicit
stack and assembly instructions to allocate, read, write, and free
stack cells. We follow much of their basic design, including
the use of stack-tail polymorphism to hide values on the stack so they will
be preserved across calls, and the use of register-file and stack
typing to specify preconditions for jumping to a code block. 

Our main novelty is identifying the notion of a TAL
\emph{component}. In $\tlang$, we need to be able to reason about a
component $\tal{e_T}$ because we will eventually be embedding these
components as terms in a high-level functional language called
$\flang$. A component $\tal{e_T}$ must be composed of assembly
instructions, but we don't want to restrict it to a single basic
block so we use a pair $(\tal{I},\tal{H})$ of an instruction sequence
$\tal{I}$ and a local heap fragment $\tal{H}$ that maps locations to
code blocks used in local intra-component jumps.

The combined language $\ftlang$ is a typical Matthews-Findler
multi-language \cite{matthews07}, where the syntax of both languages
are combined and boundary terms are added to mediate interactions
between the two. A boundary term $\bound< \SRC \TAL \tau {e_T}$ means
that the $\tlang$ component $\tal{e_T}$ within the boundary will be
used in an $\flang$ context at type $\src\tau$. To be well-typed, the
inner component $\tal{e_T}$ should have the type translated from
$\src{\tau}$ according to the multi-language type translation in
\S\ref{sec:ftlang}.

$\ftlang$ exists to enable reasoning about the equivalence of $\flang$
expressions and $\tlang$ components, or mixed combinations of the
two. Intuitively, we would like to treat blocks of assembly as similar
to functions in high-level languages. Semantically, functions are
objects that, given related inputs, produce related outputs.
Following STAL we can, at least, model the state of the stack and a
subset of the registers as inputs. But blocks of assembly instructions
do not have clear outputs to relate, leading us towards
one of our central novel contributions.

In STAL, every basic block has type $\codet \Delta \chi \sigma {}$, where
$\Delta$ contains type parameters, and $\chi$ and $\sigma$ are respectively the
register and stack typing preconditions. Since every block is in continuation
style, blocks never return, always jumping to the next block, so there never
need be outputs to relate --- the output of a block is just the input
constraints on the block to which it jumps. In our mixed-language setting we
must, therefore, provide components with return continuations which when called
from high-level code contain a halting instruction, and when called from
assembly jump to the next step in execution. In order to determine
the result type---i.e., the type of the value that is either halted
with or passed to the next block---we extend the STAL code pointer type to $\codet {\tal\Delta} {\tal\chi}
{\tal\sigma} {\tal{q}}$, where $\tal{q}$ is our critical addition.

A \emph{return marker} $\tal{q}$ specifies the register or stack position where the return
continuation is stored. This allows us, following a basic calling
convention, to determine the type of the value that will be passed to
that continuation. As we will see in later sections, there are a few
other forms that $\tal{q}$ can take, but they all support our ability
to reason about $\tlang$ components as semantic objects that produce
values of a specific type. This allows us to reason not only about the
equivalence of structurally different assembly components
made up of different numbers of basic blocks, but of components made
up of entirely different mixes of languages.


\section{Typed Assembly Language: \protect\tlang}
\label{sec:tlang}

\begin{figure}
  \begin{small}
    \begin{tabular}{rcl} 
      Value type \tal{\tau} & ::= & \tal{\alpha} $\vert$ \tal{\unitt} $\vert$ \tal{\int}
              $\vert$ \tal{\exists \alpha. \tau} $\vert$ \tal{\mu \alpha. \tau} \\
                            & & \tal{\refto {\tuplet{\tau,\ldots,\tau}}} $\vert$ \tal{\boxof \psi} \\

      Word value \tal{w} & ::= & \tal{\unit} $\vert$ \tal{n} $\vert$ \tal{\ell}
          $\vert$ \tal{\pack \tau w {\exists \alpha.\tau}} \\
                            & & \tal{\fold {\mu\alpha.\tau} w} $\vert$ \tal{w [\omega]} \\

      Register \tal{r} & ::= & \tal{\reg1} $\vert$ \tal{\reg2} $\vert$ $\cdots$ $\vert$ \tal{\reg7} $\vert$ \tal{\reg a} \\

      Small value \tal{u} & ::= & \tal{w} $\vert$ \tal{r}
          $\vert$ \tal{\pack{\tau}{u}{\exists \alpha.\tau}} \\
                            & & \tal{\fold {\mu\alpha.\tau} u} $\vert$ \tal{u [\omega]} \\
      Type instantiation \tal{\omega} & ::= & \tal{\tau}  $\vert$ \tal{\sigma} $\vert$ \tal{q} \\

      Heap value type \tal{\psi} & ::= & \tal{\codet \Delta \chi \sigma q}
                                   $\vert$ \tal{\tuplet{\tau, \ldots, \tau}} \\

      Heap value \tal{h} & ::= & \tal{\code \Delta \chi \sigma q I}
           $\vert$ \tal{\tuple{w, \ldots, w}} \\

      Register typing \tal{\chi} & ::= & \tal{\cdot} $\vert$ \tal{\chi, r\:\tau} \\

      Stack typing \tal{\sigma} & ::= & \tal{\zeta} $\vert$ \tal{\emptystackt} $\vert$ \tal{\cons \tau \sigma} \\ 

      Return marker \tal{q} & ::= & \tal{r} $\vert$ \tal{i} $\vert$ \tal{\epsilon} $\vert$ \retmark \tau \sigma \\

      Type env \tal{\Delta} & ::= & \tal{\cdot} $\vert$ \tal{\Delta, \alpha}
                $\vert$ \tal{\Delta, \zeta} $\vert$ \tal{\Delta, \epsilon} \\

      Heap typing \tal{\Psi} & ::= & \tal{\cdot} $\vert$ \tal{\Psi, \ell \:^\nu \psi} \\
      & & where $\tal{\nu} ::= \tal{\reftag} \vert \tal{\boxtag}$ \\

      Memory \tal{M} & ::= & \tal{(H, R, S)} \\ 
      
      Heap fragment \tal{H} & ::= & \tal{\cdot} $\vert$ \tal{H, \ell \mapsto h} \\

      Register file \tal{R} & ::= & \tal{\cdot} $\vert$ \tal{R, r \mapsto w} \\

      Stack \tal{S} & ::= & \tal{\emptystack} $\vert$ \tal{\cons w S} \\
      
      Instruction sequence \tal{I} & ::= & \\

      \multicolumn{3}{l}{
      \qquad \begin{tabular}{rl}
        \tal{\iota; I} & instruction sequencing \\
        \tal{\jmp u} & jump to $\tal{u}$ within same component\\
        \tal{\callinstr u \sigma q} & jump to $\tal{u}$, with return
                                      address at $\tal q$ \\
        \tal{\retinstr r {r_r}} & jump back to code at $\tal r$ with result in $\tal {r_r}$ \\
        \tal{\haltinstr \tau \sigma {r_r}} & halt with value type $\tal\tau$ in register $\tal{r_r}$ \\
      \end{tabular}
      }\\
      
      Single instruction \tal{\iota} & ::= &  \\
      
      \multicolumn{3}{l}{
      \begin{tabular}{rl}
              \tal{\binop{\aop}{r_d}{r_s}{u}} & store result of $\tal{\mathtt{add}}\vert\tal{\mathtt{mul}}\vert\tal{\mathtt{sub}}$ in $\tal{r_d}$ \\
              \tal{\bnz r u} & jump to $\tal u$ if $\tal{r}$ doesn't contain 0 \\
              \tal{\ld{r_d}{r_s}{i}} & load from $\tal i$th position in tuple
               at $\tal{r_s}$\\
              \tal{\st{r_d}{i}{r_s}} & store to $\tal i$th position in
                                       mutable tuple at $\tal{r_d}$\\
              \tal{\refallocinstr{r_d} n} & alloc mutable n-tuple
                                            from stack\\
              \tal{\boxallocinstr{r_d} n} & alloc immutable n-tuple
                                            from stack\\
              \tal{\mv{r_d}{u}} & move value $\tal u$ into register $\tal{r_d}$\\
              \tal{\salloc n} & allocate $\tal n$ stack cells with
                                unit values \\
              \tal{\sfree n} & free $\tal n$ stack cells\\
              \tal{\sld{r_d}{i}} & load $\tal i$th stack value into $\tal{r_d}$\\
              \tal{\sst{i}{r_s}} & store $\tal{r_s}$ into $\tal i$th stack slot\\
              \tal{\unpackinstr{\alpha}{r_d}{u}} & unpack existential,
                                                   binding to $\tal{\alpha,r_d}$\\
              \tal{\unfoldinstr{r_d}{u}} & unfold recursive type\\
            \end{tabular}
      } \\

      Component \tal{e} & ::= & \tal{(I, H)} \\

      Halt instruction \tal{v} & ::= & \tal{\haltinstr \tau \sigma {r_r}} \\

      Evaluation context \tal{E} & ::= & \tal{(\hole,\cdot)} \\
    \end{tabular}
    \vspace{0.25cm}
    \caption{\protect\tlang~Syntax}
    \label{tal-syntax}
  \end{small}
\end{figure}

\paragraph{Syntax} Figure~\ref{tal-syntax} presents the full syntax of
$\tlang$, our typed assembly language. Value types $\tal\tau$ are the
types ascribed to values small enough to fit in a register, including
base values, recursive and existential types, and mutable
($\tal{ref}$) or immutable ($\tal{box}$) pointers to heap values.
We ascribe value types $\tal\tau$ to word values $\tal{w}$, which 
include unit $\tal{()}$, integers $\tal{n}$, 
locations $\tal\ell$, existential $\tal{pack}$s, and recursive
$\tal{fold}$s.  We additionally follow STAL's convention that a word
value $\tal{w}$ applied to a type instantiation $\tal\omega$ is itself a
value $\tal{w[\omega]}$.  Small values $\tal{u}$ include  word
values $\tal{w}$, but also can be a register $\tal{r}$ that contains a
word value. Instructions accept small values $\tal{u}$ as operands;
hence, in the operational semantics, if $\tal{u}$ is a register we
first load the value from the register, while if $\tal{u}$ is a word
value we use it directly.

We ascribe heap-value types $\tal\psi$ to heap values $\tal{h}$.  These include tuples of word values
$\tal{\tuplet{w,\ldots,w}}$ and code blocks $\tal{\code \Delta \chi 
  \sigma q I}$, which have types $\tal{\tuplet{\tau,\ldots,\tau}}$ and 
$\tal{\codet \Delta \chi \sigma q}$, respectively. Note that we have
mutable ($\tal{ref}$) references to tuples but only immutable
($\tal{box}$) references to code, since we prohibit self-modifying
code.

Code blocks $\tal{\code \Delta \chi \sigma q I}$ specify a type
environment $\tal{\Delta}$, a register file typing $\tal\chi$, and
a stack type $\tal\sigma$ for an instruction sequence $\tal{I}$.  Here
$\tal\chi$ and $\tal\sigma$ are preconditions for safely jumping to
$\tal{I}$: $\tal\chi$ is a mapping from registers $\tal{r}$ to the  
type of values $\tal\tau$ the registers must contain, while
$\tal\sigma$ is a list of value types on top of the stack that may end
with an abstract stack-tail variable $\tal\zeta$.  The type variables
in $\tal\Delta$, which may appear free in $\tal\chi$, $\tal\sigma$,
and $\tal{I}$, must be instantiated when we jump to the code block. 
If this code block is stored at location $\tal{\ell}$, and register
$\tal{r}$ contains $\tal{\ell}$, we can jump to it via 
$\tal{\jmp {r[\vec\omega]}}$ where $\tal{\vec\omega}$ instantiates
the variables in $\tal\Delta$. (We use vector notation, e.g.,
$\tal{\vec{\omega}}$ or $\tal{\vec{\tau}}$, to denote a sequence.)

As discussed in \S\ref{sec:main-ideas}, our code blocks include a
novel return marker $\tal{q}$, which tells us where to find the
current return continuation.  Here $\tal{q}$ can be a register
$\tal{r}$ in $\tal\chi$, or a stack index $\tal{i}$ that is accessible
in $\tal{\sigma}$ (i.e., the $\tal{i}$th stack slot is not hidden in
the stack tail $\tal\zeta$). Return markers can also range over type
variables $\tal\epsilon$ which we use to abstract over return markers
(as we explain below).  There is also a special return marker
$\tal{\retmark \tau \sigma}$ which means that when the current
component finishes it should halt with a value of type $\tal\tau$ and
stack of type $\tal\sigma$.  In $\tlang$, this would mean the end of
the program with a $\tal{\mathtt{halt}}$ instruction, but within a
multi-language boundary the same $\tal{\mathtt{halt}}$ results in a
transition to the high-level language.


A memory $\tal{M}$ includes a heap $\tal{H}$ which maps
locations $\tal\ell$ to heap values $\tal{h}$, a register file
$\tal{R}$ which maps registers $\tal{r}$ to word values
$\tal{w}$, and a stack $\tal{S}$ which is a list of word values.

An instruction sequence $\tal{I}$ is a list of instructions terminated
by one of three jump instructions ($\tal{\mathtt{jmp}}$,
$\tal{\mathtt{call}}$, $\tal{\mathtt{ret}}$) or the
$\tal{\mathtt{halt}}$ instruction. The distinction between jump
instructions is a critical part of $\tlang$ explored in depth later in
this section. Individual instructions $\tal\iota$ include many
standard assembly instructions and are largely similar to STAL.

A component $\tal{e}$ is a tuple $(\tal{I}, \tal{H})$ of instructions
$\tal{I}$ and a local heap fragment $\tal{H}$. The local heap fragment
can contain multiple local blocks used by the component. We
distinguish the $\tal{\mathtt{halt}}$ instruction as a value $\tal{v}$, as it
is the only $\tlang$ instruction sequence that does not reduce.

\paragraph{Operational Semantics}
We specify a small-step operational semantics as a relation on
memories $\tal{M}$ and components $\tal{e}$:~$\config{\tal{M}}{\tal{e}} \red \config{\tal{M'}}{\tal{e'}}$. Operationally, we merge
local heap fragments to the global heap and then use the evaluation
context $\tal{E}$ to reduce instructions according to relation:~$\config{\tal{M}}{\tal{I}} \red
\config{\tal{M'}}{\tal{I'}}$. In $\tlang$, evaluation contexts $\tal{E}$ are
not particularly interesting, but when $\tlang$ is embedded within the
multi-language, $\tal{E}$ will include boundaries. While
Figure~\ref{tal-syntax} includes operational descriptions of the
instructions, the full semantics are standard and elided.

\begin{figure}
  \begin{small}
    \framebox[1.05\width][l]{$\tinstr[\TAL] \Psi \Delta \chi \sigma q {\iota} {\instrpost {\Delta'}
        {\chi'} {\sigma'} {q'}}$}  \mbox{~where~} \tal{\wfret[\TAL] \cdot \Delta \chi \sigma q}
    \vspace{-0.4cm}
    
  \begin{mathpar}
    \infer{\t[\TAL] \Psi \Delta \chi u \tau \\ \tal{q} \ne \tal{r_d} \\ \tal{u}
      \ne \tal{q}}
    {\tinstr[\TAL] \Psi \Delta \chi \sigma q {\mv{r_d}{u}}
      {\instrpost \Delta {\chi[r_d\:\tau]} \sigma q}}
    
    \and
    \infer{\tal\chi(\tal{r_s}) = \tal\tau}
    {\tinstr[\TAL] \Psi \Delta \chi \sigma {r_s} {\mv{r_d}{r_s}}
      {\instrpost \Delta {\chi[r_d\:\tau]} \sigma {r_d}}}
  \end{mathpar}

  \framebox[1.1\width][l]{$\tinseq[\TAL] \Psi \Delta \chi \sigma q I$}
  \mbox{~where~} \tal{\wfret[\TAL] \cdot \Delta \chi \sigma q}
  \vspace{-0.6cm}

  \begin{mathpar}

    \infer{\tinstr[\TAL] \Psi \Delta \chi \sigma q \iota
             {\instrpost {\Delta'} {\chi'} {\sigma'} {q'}} \\
          \tinseq[\TAL] \Psi {\Delta'} {\chi'} {\sigma'} {q'} I}
        {\tinseq[\TAL] \Psi \Delta \chi \sigma q {\iota; I}}
        \and

        \infer{\tal{\chi(r)} = \tal{\tau}}
      {\tinseq[\TAL] \Psi \Delta \chi \sigma {\retmark \tau \sigma}
                     {\haltinstr \tau \sigma r}} 
      
     \and
    
    \infer{\t[\TAL] \Psi \Delta \chi u {\boxof {\codet {} {\chi'} \sigma q}} \\
      \subt[\TAL] \Delta \chi {\chi'}}
    {\tinseq[\TAL] \Psi \Delta \chi \sigma q {\jmp u}} 

    \and
    
    \infer{\tal{\chi(r)} = \tal{\boxof {\stopt {r'} \tau \sigma {q'}}} \\
      \tal{\chi(r')} = \tal{\tau}}
    {\tinseq[\TAL] \Psi \Delta \chi \sigma r {\retinstr r {r'}}}    \and
    
    \hspace*{-0.4cm}\infer{\t[\TAL] \Psi \Delta \chi u
                {\boxof {\codet {\zeta,\epsilon}
                                {\hat\chi}
                                {\hat\sigma}
                                {\hat q}}} \\
       \wft[\TAL] \Delta {{\hat\chi}\setminus{\tal{\hat q}}} \\
       \retaddrtype(\tal{\hat q}, \tal{\hat\chi}, \tal{\hat\sigma})
          = \tal{\boxof {\stopt r \tau {\hat\sigma'} \epsilon}} \\
       \wft[\TAL] \Delta \tau \\
       \wft[\TAL] \Delta {\hat\sigma'[\sigma_0/\zeta]} \\
       \wft[\TAL] \Delta
                  {\codet {}
                          {\hat\chi [\sigma_0/\zeta]
                                    [\retmark {\tau^*} {\sigma^*}/\epsilon]}
                          {\hat\sigma [\sigma_0/\zeta]
                                      [\retmark {\tau^*} {\sigma^*}/\epsilon]}
                          {\hat q}} \\
       \subt[\TAL] \Delta \chi
                          {\hat\chi [\sigma_0/\zeta]
                                    [\retmark {\tau^*} {\sigma^*}/\epsilon]} \\\\
       \tal{\sigma} = \tal{\cons {\vec\tau} {\sigma_0}} \\
       \tal{\hat\sigma} = \tal{\cons {\vec\tau} \zeta} \\
       \tal{\hat\sigma'} = \tal{\cons {\vec{\tau'}} \zeta}}
      {\tinseq[\TAL] \Psi \Delta \chi \sigma {\retmark {\tau^*} {\sigma^*}}
                     {\callinstr u {\sigma_0} {\retmark {\tau^*} {\sigma^*}}}}

                    \and

    
    \infer{\t[\TAL] \Psi \Delta \chi u
      {\boxof {\codet {\zeta,\epsilon}
          {\hat\chi}
          {\hat\sigma}
          {\hat q}}} \\
      \wft[\TAL] \Delta {{\hat\chi}\setminus{\tal{\hat q}}} \\
      \retaddrtype(\tal{\hat q},\tal{\hat\chi},\tal{\hat\sigma})
      = \tal{\stopt r \tau {\hat\sigma'} \epsilon} \\
      \wft[\TAL] \Delta \tau \\
      \wft[\TAL] \Delta {\hat\sigma'[\sigma_0/\zeta]} \\
      \wft[\TAL] \Delta
      {\codet {}
        {\hat\chi [\sigma_0/\zeta][i{+}k{-}j/\epsilon]}
        {\hat\sigma  [\sigma_0/\zeta][i{+}k{-}j/\epsilon]}
        {\hat q}} \\
      \subt[\TAL] \Delta \chi
      {\hat\chi [\sigma_0/\zeta][i{+}k{-}j/\epsilon]} \\\\
      \tal{\sigma} = \tal{\cons {\tau_0} {\cons \cdots
          {\cons {\tau_j} {\sigma_0}}}} \\
      \tal{\hat\sigma} = \tal{\cons {\tau_0} {\cons \cdots
          {\cons {\tau_j} \zeta}}} \\
      \tal{j} < \tal{i} \\
      \tal{\hat\sigma'} = \tal{\cons {\tau_0'} {\cons \cdots
          {\cons {\tau_k'} \zeta}}}}
    {\tinseq[\TAL] \Psi \Delta \chi \sigma i {\callinstr u {\sigma_0} {i{+}k{-}j}}}

                 \end{mathpar}


  \framebox[1.08\width][l]{$\texpr[\TAL] \Psi \Delta \chi \sigma q e \tau {\sigma'}$} 
  \vspace{-0.4cm}
  
  \begin{mathpar}
    \infer{
      \tmemfrag[\TAL] \Psi H {\Psi'} \\
      \forall (\tal{\ell\:^\nu\,\psi}) \in \tal\Psi.\ \tal{\nu} = \tal{\boxtag} \\
    \typeof(\tal q, \tal\chi, \tal\sigma) = \tal{\tau}; \tal{\sigma'} \\
    \tinseq[\TAL] {(\Psi,\Psi')} \Delta \chi \sigma q I  
    }
    {\texpr[\TAL] \Psi \Delta \chi \sigma q {(I, H)} \tau {\sigma'}} 
  \end{mathpar}

  \vspace{-0.2cm}
  
    \[
  \begin{array}{l}
    \typeof(\tal{r}, \tal{\chi}, \tal{\sigma}) = \tal{\tau}; \tal{\sigma'}
    \text{if } \tal{\chi(r)} = \tal{\boxof{\stopt {r'} \tau {\sigma'} q}} \\
    \typeof(\tal{i}, \tal{\chi}, \tal{\sigma}) = \tal{\tau}; \tal{\sigma'}
    \text{if } \tal{\sigma(i)} = \tal{\boxof{\stopt {r'} \tau {\sigma'} q}} \\
    \typeof(\tal{\retmark \tau {\sigma'}}, \tal{\chi}, \tal{\sigma})
                                               = \tal{\tau}; \tal{\sigma'} \\
    \retaddrtype(\tal{r},\tal{\chi},\tal{\sigma}) =
                                                    \tal{\stopt {r'} \tau {\sigma'} {q'}} \\
    \qquad\text{if } \tal{\chi(r)} = \tal{\boxof{\stopt {r'} \tau {\sigma'} {q'}}} \\
    \retaddrtype(\tal{i},\tal{\chi},\tal{\sigma}) =
                                                    \tal{\stopt {r'} \tau {\sigma'} {q'}} \\
    \qquad\text{if } \tal{\sigma(i)} = \tal{\boxof{\stopt {r'} \tau {\sigma'} {q'}}} \\
  \end{array}
\]
    \caption{Selected \protect\tlang~Typing Rules}
    \label{tal-typing}
  \end{small}
\end{figure}

\paragraph{Type System}

In Figure~\ref{tal-typing} we present a selection of typing rules for
$\tlang$. We elide various type judgments and well-formedness
judgments for small values, heap fragments, and register files as they
are standard, focusing instead on novel rules for instructions,
instruction sequences, and components. Full details appear in our
technical appendix~\cite{patterson17:funtal-tr}.

Instructions $\tal\iota$ and instruction sequences $\tal{I}$ are typed
under a static heap $\tal\Psi$, a type environment $\tal\Delta$, a
register file typing $\tal\chi$, a stack typing $\tal\sigma$, and
return marker $\tal{q}$.  An instruction $\tal{\iota}$ may change any
of these except the static heap.
Critically, the instruction and instruction-sequence judgments impose
restrictions on the return marker $\tal{q}$ (written $\wfret[\TAL]
{\cdot} \Delta \chi \sigma q$) to ensure that a block of
instructions knows to where it is returning.  This means that $\tal{q}$
cannot be $\tal\epsilon$ and if $\tal{q}$ is a register or 
stack index its type should be visible in $\tal\chi$ or $\tal\sigma$.
The judgment $\wfret[\TAL] {\Delta'} \Delta \chi \sigma 
q$ ensures that if $\tal{q}$ is $\tal\epsilon$, then $\tal\epsilon$ is
in $\tal\Delta$ not $\tal{\Delta'}$, which in this case is
empty.\footnote{Code pointers can have $\tal\epsilon$ in the return  
marker of their return continuation, but by the time they are jumped to this
$\tal\epsilon$ must be instantiated. An example of this is shown later in this section.}
It also checks that we can look up the types expected by the return
continuation at $\tal{q}$ using $\typeof(\tal{q}, 
\tal{\chi}, \tal{\sigma})$ (see bottom of Figure~\ref{tal-typing}). 


The $\tal{\mathtt{mv}}$ instruction shown in Figure~\ref{tal-typing} has two
cases. In the first case, we are loading a small value $\tal{u}$ with type
$\tal{\tau}$ into register $\tal{\mathtt{r_d}}$, which we know is not the return
marker $\tal{q}$. We also restrict $\tal{u}$ to not be the current return
marker $\tal{q}$, as in that case the second $\tal{\mathtt{mv}}$ rule
described below must be used. After the $\tal{\mathtt{mv}}$, the
register-file typing now reflects the updated register, which we write
as $\tal{\chi[\reg d \: \tau]}$, and no other changes 
have occurred. The second case is that we are moving the value in register
$\tal{\mathtt{r_s}}$ into register $\tal{\mathtt{r_d}}$, where
$\tal{\mathtt{r_s}}$ is the current return marker so it is pointing to
the return continuation. In this case, not only do we update the
register file, we also change the return marker 
to reflect that the continuation is now in $\tal{\mathtt{r_d}}$. Other
instructions, like $\tal{\mathtt{sst}}$ and $\tal{\mathtt{sld}}$, similarly have
cases depending on whether the operation will change where the return
continuation is stored.

Instruction typing judgments are lifted to instruction sequences by
matching the postcondition of the instruction at the head of the list
to the precondition of the rest of the sequence, as shown in
Figure~\ref{tal-typing}. We illustrate how sequences are 
type-checked with the following small example.  Note that each
instruction's postcondition is used as the precondition of the next.

\begin{small}
\vspace{-3ex}
  \[
    \arraycolsep=1pt
    \begin{array}{rll}
      \tal{\cdot}~;~ \tal{\cdot}~;~ \tal{\cdot}~;~ \tal{\bullet}~;~ \tal{\reg a} \vdash &\tal{\mv{\reg 1}{42};}
      &\Rightarrow \instrpost {~\cdot~} {\reg 1\:\int~} {\bullet~} {~\reg a}\\
      
                                                                                & \tal{\salloc{1};} &\Rightarrow \instrpost
                                                         {~\cdot~} {\reg
                                                                                                      1\:\int~} {\cons\unitt\bullet~} {~\reg a}\\
                                                                                      &
                                                                                        \tal{\sst{0}{\reg 1};} &\Rightarrow \instrpost
                                                         {~\cdot~} {\reg
                                                                                                                 1\:\int~} {\cons\int\bullet~} {~\reg a}\\
    \end{array}      
  \]
\vspace{-1.5ex}
\end{small}

First, we load $\tal{42}$ into register $\tal{\reg 1}$, which is reflected in
the register file typing $\tal{\reg 1 \: \int}$. We then allocate one cell on
the stack, which starts out as $\tal{\unitt}$. Now that there is space, we can
store the value of register $\tal{\reg 1}$ into the $\tal{0}$th slot on the
stack, which is then reflected in the stack typing.

Next in Figure~\ref{tal-typing}, we show the $\tal{\mathtt{halt}}$ instruction,
which requires the $\tal{\retmark \tau \sigma}$ return marker, indicating the
type of the value in the register specified and the type of the stack. This
instruction is how $\tlang$ programs terminate; in our $\ftlang$ multi-language,
this will also be how a $\tlang$ component transfers a value back to a wrapping $\flang$
component.

Next are the three jump instructions. First is the
\emph{intra-component jump} $\tal{\mathtt{jmp}}$ instruction. This requires
that the location $\tal{u}$ being jumped to be a code pointer (of type
$\tal{\boxof {\codet {} {\chi'} \sigma q}}$) that has preconditions
$\tal{\chi'}$ and $\tal\sigma$ for the register file and stack respectively, and
return marker $\tal q$. The current register file $\tal\chi$ must be a subset of
the expected $\tal{\chi'}$, which means that we can have more registers with
values in them, but the types of registers that occur in $\tal{\chi'}$ must match.

We also, critically, require that the return marker $\tal{q}$ on the code block
being jumped to be the same as the current return marker. This captures the
intuition of an intra-component jump. As noted before, blocks being jumped to
must have fully instantiated return markers---informally, blocks cannot
abstract over their own return markers. This restriction is only on instruction
sequences; a component can have local blocks with abstract return markers.
Consider the code pointer type:

\begin{small}
\vspace{-1.5ex}
\[\tal{\boxof{\codet{\epsilon} {\reg a \: \boxof{\codet{} {\reg1\:\tal\tau}
          \sigma \epsilon}} {\sigma} {\reg a}}}\]
\vspace{-2.5ex}
\end{small}

\noindent This type is a pointer to a code block with a return marker type
parameter $\tal\epsilon$ that requires a stack of type $\tal\sigma$
and for register $\tal{\reg a}$, the return marker, to be a code pointer. This inner code
pointer is the continuation, as the entire block has $\tal{\reg a}$ as
its return marker, but the return marker for this continuation is
$\tal\epsilon$. When the continuation in $\tal{\reg a}$ is jumped to
it requires that the stack still have type $\tal\sigma$ and that a
value of type $\tal\tau$ be stored in register $\tal{\reg 1}$. Since
code pointers can't be jumped to until all their type variables are
instantiated, the caller of this whole code block must provide a
concrete continuation in register $\tal{\reg a}$ and instantiate
$\tal\epsilon$ with the corresponding concrete return marker before jumping.

As a concrete example consider the following well-typed $\tal{\mathtt{jmp}}$
instruction:

\begin{small}
\vspace{-3ex}
  \[
    \begin{array}{l}
      \tal{\ell \:^{\boxtag} \codet{} {\reg 2 \: \unitt} {\cons\int\bullet}
        {\retmark{\unitt}{\bullet}}}~;~ \tal{\cdot}~; \\
      \quad \tal{\reg 1 \: \int, \reg 2 \: \unitt}~;~ \tal{\cons\int\bullet}~;~ \tal{\retmark{\unitt}{\bullet}} \vdash
      \tal{\jmp{\ell}}\\
    \end{array}
  \]
\vspace{-2ex}
\end{small}

As required, the $\tal{\mathtt{jmp}}$ is to a code block $\tal\ell$
that has the same return marker $\tal{\retmark{\unitt}{\bullet}}$. The
current registers has $\tal{\reg 1}$ set, which the block does not
require, but also has the register $\tal{\reg 2}$ set that the block
does require. Finally, the stack type $\tal{\cons\int\bullet}$ matches
what the block expects. Note that since the stack currently has an
$\tal{\int}$ on it but the return marker says the stack must be empty,
we will have to pop the integer off the stack either in the block
$\tal\ell$ or in some subsequent block that we jump to from
$\tal\ell$ before we $\tal{\mathtt{halt}}$.

The next instruction in Figure~\ref{tal-typing} is $\tal{\mathtt{ret}}$, which
is the \emph{inter-component} jump for returning from a
component. Notably, the location being jumped to must be in a
register; if it were still on the stack the type of $\tal\sigma$
would include itself.  We require, first, that the register $\tal{\mathtt{r}}$
being jumped to points to a code block with no type variables, and
second that the register $\tal{\mathtt{r'}}$ map to type $\tal\tau$, as required
by the block being returned to. This is a type-enforced calling
convention for the return value. Importantly, we make no restriction on the return
marker $\tal{q'}$ on the block being jumped to. This is because with
$\tal{\mathtt{ret}}$ we are jumping back to a different component,
which will in turn have its own return marker.

The last two typing rules shown in Figure~\ref{tal-typing} are for the
$\tal{\mathtt{call}}$ instruction which is our other
\emph{inter-component} jump.  The first applies when the current
component will terminate by $\tal{\mathtt{halt}}$ing.  The second
applies when the current component will terminate by jumping to
another $\tlang$ component. 

In some assembly languages, there is a convention that certain
registers (``callee-saved'') will be preserved such that when a
$\tal{\mathtt{call}}$ returns, those registers have the same values as
before. However, we follow STAL in protecting values solely through
stack-tail polymorphism, where a value can be stored in a part of the
stack that has been abstracted away as a type variable. Static typing
ensures that a callee that tried to read, write, or free values within
the abstract tail would not type check. Values that are accessible can
be passed in front of the abstract tail, and the callee is free to
allocate values in front, but typing constraints may force them to
free the values before returning.

As a concrete example of the first typing rule, consider the following
well-typed $\tal{\mathtt{call}}$ instruction:

\begin{small}
\vspace{-3ex}
  \[
    \begin{array}{l}
      \tal{\ell \:^{\boxtag} \codet{\zeta,\epsilon} {\reg
      a \: \boxof{\codet{} {\reg 1 \: \int} \zeta \epsilon}}
      {\cons\unitt\zeta}
        {\reg a}}~;~ \tal{\cdot}~; \\
      \tal{\reg 1 \: \int, \reg a \: \boxof{\codet{}{\reg 1 \: \int}{\cons\int\bullet}{\retmark{\int}{\bullet}}}}~;\\
      \tal{\cons{\unitt}{\cons\int\bullet}}~;~ \tal{\retmark{\unitt}{\bullet}} \\
\qquad \vdash \tal{\call{\ell}{\cons\int\bullet}{\retmark{\int}{\bullet}}}\\
    \end{array}
  \]
\vspace{-2ex}
\end{small}

We focus here on the stack and return continuation. The
$\tal{\mathtt{call}}$ instruction specifies a tail
$\tal{\cons\int\bullet}$ to protect.  The block at $\tal{\ell}$ being
jumped to must have a stack that has the same front and an abstract
tail, here $\tal{\cons\unitt\zeta}$. Further, the block being jumped to
must return to a continuation (here stored at $\tal{\reg a}$) with an
abstract return marker $\tal\epsilon$.  Once $\tal\epsilon$ is
instantiated with $\tal{\retmark{\int}{\bullet}}$ the return
continuation must match the current register file typing.

In the second $\tal{\mathtt{call}}$ typing rule, the return
marker is a stack position $\tal{i}$. The index $\tal{i}$ must be
greater than the number of entries $\tal{j}$ on the input stack
$\tal\sigma$ in front of the tail $\tal{\sigma_0}$ specified in the
instruction. The location being jumped to, $\tal{u}$, must be a code
pointer with input registers 
$\tal{\hat\chi}$ and stack $\tal{\hat\sigma}$. Note that the prefix of
$\tal{\hat\sigma}$ matches the prefix of $\tal\sigma$, $\tal{\cons {\tau_0}
  {\cons \cdots {\tau_j}}}$, but $\tal{\hat\sigma}$ has the abstract tail
$\tal\zeta$.

The final formal parameter to $\tal{\mathtt{call}}$, $\tal{i+k-j}$, is the return
marker that the continuation for $\tal{u}$ must use. In particular,
this is computed by taking the starting stack position $\tal{i}$ and
then noting how the stack is modified between the input stack
$\tal{\hat\sigma}$ and output stack $\tal{\hat\sigma'}$ by the code
block pointed to by $\tal{u}$. After the call, the stack has $\tal{k}$
values in front but we know that position $\tal{i}$ was beyond the
exposed $\tal{j}$ values, so the value on the stack at position
$\tal{i}$ is now at position $\tal{i+k-j}$.

The fact that $\retaddrtype(\tal{\hat
  q},\tal{\hat\chi},\tal{\hat\sigma})$ is 
$\tal{\stopt r \tau {\hat\sigma'} \epsilon}$ ensures that the block
being jumped to has a return continuation where a value of type
$\tal\tau$ is stored in some register, the stack has type
$\tal{\hat\sigma'}$, and the return marker is
$\tal\epsilon$. Operationally $\tal{u}$ will get instantiated with
$\tal{i+k-j}$ for $\tal\epsilon$, which based on the form of
$\tal{\hat\sigma'}$ means that the return continuation has preserved
the original return location.

The register file subtyping constraint

\begin{small}
\vspace{-1ex}
\[\subt[\TAL] \Delta \chi {\hat\chi
    [\sigma_0/\zeta][i{+}k{-}j/\epsilon]}\]
\vspace{-2.5ex}
\end{small}

\noindent ensures that the current register type $\tal\chi$ is a subtype of the
target $\tal{\hat\chi}$ once it has been concretely instantiated with
the stack tail and return address.

We similarly check with

\begin{small}
\vspace{-3.5ex}
\[\wft[\TAL] \Delta {\codet {} {\hat\chi
      [\sigma_0/\zeta][i{+}k{-}j/\epsilon]} {\hat\sigma
      [\sigma_0/\zeta][i{+}k{-}j/\epsilon]} {\hat q}}\]
\vspace{-3ex}
\end{small}

\noindent that the code block type is well-formed when concretely instantiated,
and with $\wft[\TAL] \Delta {\hat\sigma'[\sigma_0/\zeta]}$ that the
resulting stack is well-formed once concretely instantiated. Finally,
we ensure with $\wft[\TAL] \Delta {{\hat\chi}\setminus{\tal{\hat q}}}$
that if $\tal{\hat{q}}$ is a register then $\tal{\hat\chi}$ is well-formed
without it.  This means that while $\tal{\hat{q}}$ may have free type
variables $\tal\epsilon$ and $\tal\zeta$, the rest of $\tal{\hat\chi}$
cannot.

\begin{figure}
  \begin{small}
    \renewcommand{\arraystretch}{\sfactor}
    \[
      \begin{array}{l}
        \tal{f} = \tal{(\mv {\reg a} {\ell_{1ret}}; \callinstr{\ell_1}{\bullet}{\retmark \int \bullet}, \tal{H})} \\ 
        \tal{H(\ell_1)} =
        \begin{stackTL}
                  \tal{\code { \zeta, \epsilon}
                    {\reg a \:
                      \stopt {\reg1}
                      {\tal\int}
                      \zeta
                      \epsilon}
                    \zeta
                    {\reg a} {}}\\
                  \quad \tal{\salloc 1; \sst 0 {\reg a}; \mv {\reg a} {\ell_{2ret}[\zeta,\epsilon]};} \\
                  \quad \tal{\callinstr {\ell_2} {\cons{\stopt {\reg1}
                      {\tal\int}
                      \zeta
                      \epsilon}\zeta} {0}}
                \end{stackTL} \\
        \tal{H(\ell_{1ret})} = 
          \begin{stackTL}
                  \tal{\code{}
                    {\reg 1 \: \int}
                    \bullet
                    {\retmark \int \bullet}{}}\\
                  \quad   \tal{\haltinstr \int \bullet {\reg 1}}
                \end{stackTL} \\        
        \tal{H(\ell_2)} = 
         \begin{stackTL}
                  \tal{\code { \zeta, \epsilon}
                    {\reg a \:
                      \stopt {\reg1}
                      {\tal\int}
                      \zeta
                      \epsilon}
                    \zeta
                    {\reg a} {}}\\
                  \quad  \tal{\mv {\reg 1} {1}; \jmp {\ell_{2aux}[\zeta,\epsilon]}}
                \end{stackTL}\\
        \tal{H(\ell_{2aux})} = 
         \begin{stackTL}
                  \tal{\code { \zeta, \epsilon}
                    {\reg 1 \: {\tal\int}, \reg a \:
                      \stopt {\reg1}
                      {\tal\int}
                      \zeta
                      \epsilon}
                    \zeta
                    {\reg a} {}}\\
                  \quad  \tal{\binop\mult{\reg 1}{\reg 1} 2; \retinstr {\reg a} {\reg 1}}
                \end{stackTL}\\
        \tal{H(\ell_{2ret})} = 
         \begin{stackTL}
                  \tal{\code{\zeta,\epsilon}
                    {\reg 1 \: \int}
                    {\cons{\codet{}
                    {\reg 1 \: \int}
                    \zeta
                    {\epsilon}}{\zeta}}
                    0 {}}\\
                  \quad   \tal{\sld {\reg a} 0; \sfree 1; \retinstr {\reg a} {\reg 1}}
                \end{stackTL} \\
      \end{array}
    \]
    \vspace{-0.25cm}
    \caption{\protect\tlang~Example: Call to Call}
    \label{call-to-call}
  \end{small}
  \vspace{-0.25cm}
\end{figure}

\begin{figure}
  \begin{small}

    \begin{tikzpicture}
            
      \node (f) [talblock] at (0,2) {$\tal{f}$};
      \node (l1) [talblock] at (3,2) {$\tal{\ell_1}$};
      \node (l2) [talblock] at (6,2) {$\tal{\ell_2}$};
      \node (l2aux) [talblock] at (6,-1) {$\tal{\ell_{2aux}}$};
      
      \node (l2ret) [talblock] at (3,0.5) {$\tal{\ell_{2ret}}$};
      \node (l1ret) [talblock] at (0,0.5) {$\tal{\ell_{1ret}}$};
      \node (halt) [vertex] at (0,-1.5) {}; 

      \draw [arrow] (f) --
        node[memory] {\calledge{call}{\reg a \map \ell_{1ret}}{\bullet}}
        (l1);

      \draw [arrow] (l1) --
        node[memory] {\calledge{call}{\reg a \map \ell_{2ret}}{\cns{\ell_{1ret}}\bullet}}
        (l2);

      \draw [arrow] (l2) --
        node[memory] {\calledge{jmp}{\reg 1 \map 1, \reg a \map \ell_{2ret}}{\cns{\ell_{1ret}}\bullet}}
        (l2aux);

      \draw [arrow] (l2aux) to [out=180, in=270]
        node[memory, pos=0.4] {\calledge{ret}{\reg 1 \map 2}{\cns{\ell_{1ret}}\bullet}}
        (l2ret);
        
      \draw [arrow] (l2ret) --
        node[memory] {\calledge{ret}{\reg 1 \map 2}{\bullet}}
        (l1ret);

      \draw [arrow] (l1ret) --
        node[memory] {\calledge{halt}{\reg 1 \map 2}{\bullet}}
        (halt); 
      
    \end{tikzpicture}
    
    \caption{\protect\tlang~Control Flow: Call to Call (Fig.~\ref{call-to-call})}
    \label{control-call-to-call}
  \end{small}
  \vspace{-0.25cm}
\end{figure}
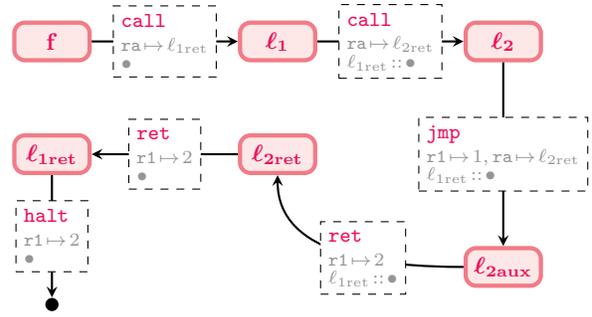

\paragraph{Example}In Figure~\ref{call-to-call}, we show an example $\tlang$ program
demonstrating
$\tal{\mathtt{call}}$, $\tal{\mathtt{jmp}}$, $\tal{\mathtt{ret}}$, and
$\tal{\mathtt{halt}}$. The control flow, in
Figure~\ref{control-call-to-call}, shows the instructions causing
jumps between basic blocks and the state of the relevant
registers and stack at jump-time. In this diagram, $\tal{\ell_2}$ and
$\tal{\ell_{2aux}}$ are in the same component, while the rest are made
up of distinct components that together make up the component $\tal{f}$.


\section{\protect\ftlang~Multi-Language}
\label{sec:ftlang}

We present a minimal functional language $\flang$ and then embed $\flang$
and $\tlang$ within a Matthews-Findler style
multi-language. Particularly notable are the boundary translations for
higher-order functions and code blocks. In \S\ref{sec:lr}, we
design a logical relation with which we can show equivalence of
programs that differ both structurally and algorithmically.

\begin{figure}[h]
  \begin{small}
   \vspace{-0.25cm}
    \[
      \hspace*{-0.5cm}
      \begin{tabular}{rcl}
      Type \src{\tau} & ::= & \src{\alpha} $\vert$ \src{\unitt} $\vert$ \src{\int}
             $\vert$ \src{\arrow{\vec{\tau}}{\tau}}
             $\vert$ \src{\mu \alpha. \tau}
             $\vert$ \src{\tuplet{\vec\tau}} \\
        Expression \src{e} & ::= & \src{x} $\vert$ \src{\unit} $\vert$ \src{n} $\vert$ \src{e\ p\ e}
          $\vert$ \src{\test e e e} $\vert$ \src{\func{\vec{x\:\tau}}{e}}
                                   $\vert$ \src{\sapp{e}{\vec{e}}} \\
        & & \src{\fold {\mu\alpha.\tau} e} $\vert$ \src{\unfold e}
            $\vert$ \src{\tuple{\vec e}} $\vert$ \src{\proj i e} \\
                      & & where \src{p} ::= \src{+} $\vert$ \src{-} $\vert$ \src{*} \\
        Value \src{v} & ::= & \src{\unit} $\vert$ \src{n} $\vert$ \src{\sfunc{\vec{x\:\tau}}{e}}
          $\vert$ \src{\fold {\mu\alpha.\tau} v} $\vert$ \src{\tuple{\vec v}} \\
        Evaluation ctxt \src{E} & ::= & \src{\hole}
          $\vert$ \src{E\ p\ e} $\vert$ \src{v\ p\ E}
          $\vert$ \src{\test {E} e e} $\vert$ \src{\sapp{E}{\vec{e}}}
          $\vert$ \src{\sapp{v}{\vec{v}\, E\, \vec{e}}} \\
          & & \src{\fold {\mu\alpha.\tau} {E}} $\vert$ \src{\unfold {E}}
          $\vert$ \src{\tuple{\vec v, E, \vec e}} $\vert$ \src{\proj i {E}} \\
      \end{tabular}
    \]
    \caption{\protect\flang~Syntax}
    \label{f-syntax}
  \end{small}
 \vspace{-0.25cm}
\end{figure}

\subsection{Functional Language: \protect\flang}

In Figure~\ref{f-syntax} we present the syntax of $\flang$, our
simply-typed call-by-value functional language with iso-recursive
types, conditional branching, tuples, and base value integers and
unit. The language is featureful enough to implement simple programs,
while lacking certain expressiveness (like mutation) that we can add
by way of the embedded assembly. The typing and operational semantics
are standard and provided in the technical appendix \cite{patterson17:funtal-tr}.

\subsection{Embedding \protect\tlang~in \protect\ftlang}

\paragraph{Syntax} In Figure~\ref{ft-syntax} we present the syntax of
our multi-language $\ftlang$, which is largely made up of extensions
to syntactic categories of either $\tlang$ (Figure~\ref{tal-syntax})
or $\flang$ (Figure~\ref{f-syntax}). Note that both expressions
$\src{e}$ and components $\tal{e}$ are components $e$ in this
language. Henceforth, when we refer to an $\flang$ or $\tlang$ term we
are referring to the terms that originated in that language, which can
now of course include nested components of the other language. We add
boundaries $\bound<\SRC\TAL\tau e$ ($\tlang$ inside, $\flang$ outside)
and $\bound>\TAL\SRC\tau e$ ($\flang$ inside, $\tlang$ outside) to
mediate between the languages. In both cases, the $\flang$ type
$\src\tau$ directs the translation. In particular, the
$\bound<\SRC\TAL\tau e$ contains a $\tlang$ component $\tal{e}$ with
$\tlang$ translated type $\tytrans\SRC\TAL\tau$, while the
$\bound>\TAL\SRC\tau e$ contains an $\flang$ expression $\src{e}$ of
type $\src\tau$. Like Matthews-Findler \cite{matthews07}, we reduce
the component within the boundary to a value, after which we carry out
a type-directed value translation using translation metafunctions
$\transval<\SRC\TAL \tau \cdot$ and $\transval>\TAL\SRC \tau \cdot$, e.g.:

\vspace{-2ex}
\[
\bound<\SRC\TAL\tau e \mathrel{\red[*]} \bound<\SRC\TAL\tau v \mathrel{\red} \transval<\SRC\TAL \tau v
\]
\vspace{-3.5ex}

\begin{figure}
  \begin{small}
   \vspace{-0.5cm}
    \[
      \begin{tabular}{rcl}
        Type \src{\tau} & ::= & $\cdots$ $\vert$ {\src{\sarrow{\vec\tau}{\pref}{\pref}{\tau'}}} \\
        Expression \src{e} & ::= & $\cdots$ $\vert$  $\bound< \SRC \TAL \tau e$ $\vert$ \src{\sfuncz{\tal{\pref}}{\tal{\pref}}{\vec{x\:\tau}}{t}}  $\vert$ \src{\app{t}{\vec{t'}}} \\ 
        Return marker \tal{q} & ::= & $\cdots$ $\vert$ \externmark \\
        Instruction sequence \tal{I} & ::= & $\cdots$ $\vert$ \tal{\protec{\pref}{\zeta}; I} \\
        Instruction \tal{\iota} & ::= & $\cdots$ $\vert$ \tal{\import{r_d}{\sigma}{\bound>\TAL\SRC\tau e}} \\
        Stack prefix \tal{\pref} & ::= & \tal{\cdot} $\vert$ \tal{\cons \tau {\pref}} \\
        Stack typing \tal{\sigma} & ::= & \tal{\cons \pref \zeta} $\vert$ \tal{\cons \pref \emptystackt} \\
        Evaluation ctxt \src{E} & ::= & $\cdots$ $\vert$ $\bound< \SRC \TAL \tau E$ \\
        Evaluation ctxt \tal{E} & ::= & $\cdots$ $\vert$ $(\tal{\import{r_d}{\sigma}{\bound>\TAL\SRC\tau E}; I}, \cdot)$ \\
        \hline
        Type $\tau$ & ::= & \src \tau $\vert$ \tal \tau \\
        Component $e$ & ::= & \src e $\vert$ \tal e \\
        $\Delta$ & ::= & $\cdot$ $\vert$ $\Delta$, \src{\alpha} $\vert$ $\Delta$, \tal{\alpha} $\vert$ $\Delta$, \tal{\zeta}
                         $\vert$ $\Delta$, \tal{\epsilon} \\
        Evaluation ctxt E & ::= & \src E $\vert$ \tal E \\ 
      \end{tabular}
    \]
    \caption{\protect\ftlang~Multi-Language Syntax}
    \label{ft-syntax}
  \end{small}
  \vspace{-0.25cm}
\end{figure}

To $\tlang$ instructions $\tal\iota$, we add an
$\tal{\mathtt{import}}$ instruction to wrap the boundary and to
specify what register the translated value should be placed in. The
$\tal{\mathtt{import}}$ instruction also specifies $\tal\sigma$, the
tail of the stack that should be protected while evaluating the
$\flang$ expression $\src{e}$, which could in turn include $\tlang$
code. Consider the following concrete example, which computes the $\flang$
expression $\small{\src{1 + 1}}$ and loads it into register $\tal{\reg 1}$, protecting the
whole stack---here, just the empty stack---while doing it:

\begin{small}
\vspace{-3ex}
  \[
    \begin{array}{l}
      \tal{\cdot}~;~ \tal{\cdot}~;~ \tal{\cdot}~;~ \tal{\bullet}~;~
      \tal{\retmark\int\bullet} \vdash
      \tal{\import{\reg
      1}{\bullet}{\bound>\TAL\SRC\int{(1+1)}}} \\
      \qquad \Rightarrow \instrpost {~\cdot~} {\reg 1\:\int~} {\bullet~} {~\retmark\int\bullet}
    \end{array}
  \]
\vspace{-2ex}
\end{small}

When translating $\tlang$ code blocks into $\flang$ functions, we will
need to instantiate the stack tail variable $\tal\zeta$ on the
$\tlang$ code block. For this reason, we introduce the
$\tal{\mathtt{protect}}$ instruction, which specifies a stack prefix
$\tal{\pref}$ to leave visible and a type variable $\tal\zeta$ to bind
to the tail. We will see the value translation later in the section.

While normal $\flang$ lambdas are embedded in the multi-language, they do not
allow stack modification in embedded $\tlang$ code. However, we may want to
allow that sort of modification. For this reason, we introduce an optional new
stack-modifying lambda term
$\src{\sfuncz{\tal{\pref_i}}{\tal{\pref_o}}{\vec{x\:\tau}}{e}}$, which specifies
the stack prefix $\tal{\pref_i}$ it requires on the front of the stack when it
is called, and the stack prefix $\tal{\pref_o}$ that it will have replaced
$\tal{\pref_i}$ with upon return. Correspondingly, we introduce a new arrow type
that captures that relationship. Note that the ordinary lambda can be seen as a
special case when $\tal{\pref_i}$ and $\tal{\pref_o}$ are both the empty prefix
$\tal\cdot$, which corresponds to the entire stack being the protected
tail.  While there is no fundamental reason these stack-modifying
lambdas must be included, we can use them, for instance, to write a
function that pushes the number $\tal{7}$ onto the stack using
embedded assembly:

\begin{small}
\vspace{-3ex}
  \[
      \src{\sfuncz{\tal{\bullet}}{\tal{\cons\int\bullet}}{\vec{x\:\int}}{}}
      \bound< \SRC \TAL \unitt {}\tal{(}\begin{stackTL}
                                         \tal{\protec{\cdot}{\zeta}; \mv{\reg 1}{7}; \salloc 1;}\\
                                         \tal{\sst 0 {\reg 1}; \mv{\reg 1}{\unit};}\\
                                         \tal{\haltinstr{\unitt}{\cons{\int}{\zeta}}{\reg
                                         1},\cdot)}
                                         \end{stackTL} 
  \]
\vspace{-2ex}
\end{small}

The inline assembly of this function first captures the current stack
as an abstract $\tal\zeta$, then loads $\tal{7}$ into register $\tal{\reg 1}$,
allocates a cell on the stack and stores the value there, before clearing out
$\tal{\reg 1}$ and halting on it. Without stack-modifying lambdas, this would
fail to type check since the stack at the end of the body of the lambda would
be different than it had been at the beginning. In our technical appendix and artifact we
use this feature to implement a very basic mutable reference library.

Finally, in Figure~\ref{ft-syntax} we also add a new return marker
$\tal\externmark$, which is used for 
$\flang$ code, since $\flang$ follows normal expression-based evaluation and
thus has no return continuation. 

\begin{figure}
  \begin{small}
    \framebox[1.08\width][l]{\tfull {\tal \Psi} {\tal{\Delta}; {\src \Gamma}} \chi \sigma q e \tau {\sigma'}}
    \vspace{-0.2cm}

    \begin{mathpar}

      \infer{{\tfull {\tal \Psi} {\tal{\Delta}; {\src{\Gamma}}} \chi {\sigma} \externmark
          {\src{t}}{\src{\arrow{\tau_1 \cdots
                \tau_n}{\tau'}}} {\sigma_0}}\\
        {\tfull {\tal \Psi} {\tal{\Delta}; {\src{\Gamma}}} \chi {\sigma_{i-1}} \externmark
          {\src{ t_i}}{\src{{\tau_i}}} {\sigma_i}}}
      {{\tfull {\tal \Psi} {\tal{\Delta}; {\src{\Gamma}}} \chi {\sigma} \externmark
          {\src{\app{t}{t_1\cdots t_n}}}{\src{\tau'}} {\sigma_n}}} 

      \and
      
      \infer{\tfull {\tal{\Psi}} {\tal{\Delta}; \src{\Gamma}} \cdot \sigma
        {\retmark {\tytrans\SRC\TAL \tau} {\sigma'}} {\tal{e}}
        {\tytrans\SRC\TAL \tau} {\sigma'}}
      {\tfull {\tal{\Psi}} {\tal{\Delta}; \src{\Gamma}} \chi \sigma \externmark
        {\bound<\SRC\TAL{\tau} e} {\src \tau} {\sigma'}}

      \and
      
      \infer{{\tfull {\tal \Psi} {\tal{\Delta, \zeta}; {\src{\Gamma},\src{\vec{x\:\tau}}}}
          \chi {\cons {\pref_i} {\tal{\zeta}}} \externmark
          {\src t} {\src {\tau'}} {\cons {\pref_o} \zeta}}}
      {{\tfull {\tal \Psi} {\tal{\Delta}; {\src \Gamma}} \chi
          {\sigma} \externmark
          {\src{\sfuncz{\tal{\pref_i}}{\pref_o}{\vec{x\:\tau}}{t}}}
          {\src{\sarrow{\vec\tau}{\pref_i}{\pref_o}{\tau'}}}
          {\sigma}}}

    \end{mathpar}
    \framebox[1.1\width][l]{\tinseq {\tal \Psi} {\tal{\Delta}; {\src \Gamma}} \chi \sigma q I}
      \mbox{~where~} \tal{\wfret[\TAL] \cdot \Delta \chi \sigma q}
      \vspace{-0.2cm}

    \begin{mathpar}
   \infer{\tal{\sigma} = \tal{\cons{\pref}
                               {\sigma_0}} \\
          \tal{\sigma'} = \tal{\cons{\pref}
            {\zeta}} \\
          {\tinseq {\tal\Psi} {\tal{\Delta, \zeta}; {\src \Gamma}}
            {\tal{\chi}} {\tal{\sigma'}} {q} I}}
        {\tinseq {\tal \Psi} {\tal{\Delta}; {\src \Gamma}} \chi \sigma {q}
          {\tal{\protec {\pref} {\zeta}}; I}}
      \end{mathpar}

    \framebox[1.05\width][l]{\tinstr {\tal \Psi} {\tal{\Delta}; {\src \Gamma}} \chi \sigma q {\iota}
      {\instrpost[]{\tal{\Delta'}}{\chi'}{\sigma'}{q'}}}
  \mbox{~where~} \tal{\wfret[\TAL] \cdot \Delta \chi \sigma q}
    \vspace{-0.6cm}

    \begin{mathpar}
      
      \hspace*{-0.75cm}\infer{\tal{\sigma} = \tal{\cons{\tau_0}
                               {\cons\cdots
                                 {\cons{\tau_j}{\sigma_0}}}} \\
          \tal{\sigma'} = \tal{\cons{\tau_0'}
                                {\cons\cdots
                                  {\cons{\tau_k'}{\sigma_0}}}} \\
                             \tal{\sigma^*} = \tal{\cons{\tau_0}
                               {\cons\cdots
                                 {\cons{\tau_j}{\zeta}}}} \\
          \tal{\sigma'^*} = \tal{\cons{\tau_0'}
                                {\cons\cdots
                                  {\cons{\tau_k'}{\zeta}}}} \\ 
                                  \tfull {\tal{\Psi}} {{\tal{\Delta},\tal\zeta}; {\src \Gamma}} \chi
                 {\sigma^*}
                 \externmark {\src e} {\src \tau}
                 {\sigma'^*} \\
          \tal{q} = \tal{i}>\tal{j} \text{ or }
          \tal{q} = \tal{\retmark {\hat\tau} {\hat\sigma}}}
        {\tinstrbreak {\tal \Psi} {\tal{\Delta}; {\src \Gamma}} \chi \sigma {q}
            {\tal{\import {r_d} {\sigma_0} {\bound>\TAL\SRC{\tau} e}}}
            {\instrpost[] {\tal{\Delta}} {\tal{(r_d \: \trans\SRC\TAL \tau)}}
              {\tal{\sigma'}} {\incq q {k{-}j}}}}
      
    \end{mathpar}
    \caption{Selected \protect\ftlang~Typing Rules}
    \label{ft-typing}
  \end{small}
 \vspace{-0.25cm}
\end{figure}

\paragraph{Type System}

The typing judgments for $\ftlang$, for which we show a selection in
Figure~\ref{ft-typing}, include modified versions from both $\tlang$
and $\flang$ judgments as well as rules for the new forms. Since this
is a multi-language and not a compiler, the typing rules for $\tlang$
must now include an $\flang$ environment $\src\Gamma$ of free $\flang$
variables. Similarly, the typing rules for $\flang$ must now include
all of the context needed by $\tlang$, since in order to type-check
embedded assembly components we will need to know the current register
($\tal\chi$), stack ($\tal\sigma$), and heap ($\tal\Psi$) typings.

Most of these modifications are straightforward; we show the rule for
$\flang$ application in Figure~\ref{ft-typing} as a
representative. Note that the stack typings $\tal{\sigma_i}$ are
threaded through the arguments according to evaluation order, as each
one could include embedded $\tlang$ code that modified the stack.

\begin{figure}
  \begin{small}
    \[\begin{stackTL}
      \config {\tal M} {E[\bound<\SRC\TAL \tau
        {\tal{(\haltinstr {\tytrans\SRC\TAL \tau} \sigma r,\cdot)}}]} \\
\quad \red[] \config {\tal {M'}} {E[\src{v}]}
        \hfill \text{if } \transval<\SRC\TAL \tau {M.R(r) \withmem {\tal M}} = (\src{v},\tal{M'}) \\
               \config {\tal M} {E[{\tal{\import{r_d}{\sigma'}{\bound>\TAL\SRC \tau v}; I}}]} \\
\quad \red[] \config {\tal{M'}} {E[\tal{\mv {r_d} w; I}]} \hspace{2em}\hfill \text{if } \transval>\TAL\SRC \tau {v \withmem {\tal{M}}} = (\tal{w},\tal{M'})
\end{stackTL}\]
\vspace{-0.1cm}
    \caption{\protect\ftlang~Operational Semantics: Language Boundaries}
    \label{ft-boundary-semantics}
  \end{small}
 \vspace{-0.5cm}
\end{figure}

\begin{figure}[h!]
  \begin{small}
    \[
      \begin{stackTL}
        \begin{aligned}
          \tytrans\SRC\TAL \alpha &= {\tal \alpha} \\
          \tytrans\SRC\TAL \unitt &= \tal\unitt &
          \tytrans\SRC\TAL{\mu\alpha.\tau} &=
          \tal{\mu\alpha.(\tytrans\SRC\TAL \tau)} \\
          \tytrans\SRC\TAL \int &= \tal\int &\quad
          \tytrans\SRC\TAL{\tuplet{\tau_1,\ldots,\tau_n}} &=
          \tal{\boxof{\tuplet{\tytrans\SRC\TAL{\tau_1}, \ldots,
                \tytrans\SRC\TAL{\tau_n}}}} \\
        \end{aligned} \\

        \tytrans\SRC\TAL
{\arrow {\tau_1,\ldots,\tau_n}{\tau'}} = \\
\quad \tal{\boxof{\codet{\zeta,\epsilon}
                  {\reg a \:
                     \boxof{\codet{}
                            {\reg1\:\tytrans\SRC\TAL{\tau'}}
                            \zeta \epsilon}}
                        {\sigma'} {\reg a}}} \\
                    \hspace{2.4cm} \text{where } {\tal{\sigma'}} =
 \tal{\cons{\tytrans\SRC\TAL{\tau_n}}
           {\cons{\cdots}
                 {\cons{\tytrans\SRC\TAL{\tau_1}}
                   \zeta}}} \\

        \tytrans\SRC\TAL
{\sarrow {\tau_1,\ldots,\tau_n}{\pref_i}{\pref_o}{\tau'}} = \\
\quad \tal{\boxof{\codet{\zeta,\epsilon}
                  {\reg a \:
                     \boxof{\codet{}
                            {\reg1\:\tytrans\SRC\TAL{\tau'}}
                            {\cons {\pref_o} \zeta} \epsilon}}
                        {\sigma'} {\reg a}}} \\
                    \hspace{2.4cm} \text{where } {\tal{\sigma'}} =
 \tal{\cons{\tytrans\SRC\TAL{\tau_n}}
           {\cons{\cdots}
                 {\cons{\tytrans\SRC\TAL{\tau_1}}
                   {\cons {\pref_i} \zeta}}}}
           \end{stackTL}
         \]
         \vspace{-0.1cm}
    \caption{\protect\ftlang~Boundary Type Translation}
    \label{ft-type-translation}
  \end{small}
  \vspace{-0.25cm}
\end{figure}

For the boundary term, $\bound<\SRC\TAL \tau e$, we require that the
$\tlang$ component $\tal{e}$ within the boundary be well typed under
translation type $\tytrans\SRC\TAL\tau$ and return marker
$\tal{\retmark {\tytrans\SRC\TAL\tau} {\sigma'}}$, which corresponds
to the inner assembly halting with a value of type
$\tytrans\SRC\TAL\tau$. In that case, the boundary term is well typed
under $\src\tau$ at the $\externmark$ return marker that corresponds
to $\flang$ code. Note that the boundary makes no restriction on
modification of the stack. Also in the figure is the typing rule for
the stack-modifying lambda term, which is an ordinary lambda typing
rule except it types under stacks with the given
prefixes $\tal{\pref_i}$ and $\tal{\pref_o}$ and abstract tails
$\tal\zeta$; as noted before, the regular lambda is a special case when
$\tal{\pref_i}$ and $\tal{\pref_o}$ are empty.

As described above, we add two new $\tlang$ instructions. The
$\tal{\mathtt{protect}}$ instruction is used to abstract the tail of the stack,
which we can see in the transformation of the stack
$\tal{\cons{\pref}{\sigma_0}}$ into
$\tal{\cons{\pref}{\zeta}}$ when typing the subsequent
instruction sequence $\tal{I}$, where $\tal\zeta$ is a new type variable
introduced to the type env\-ironment.  Note that there is no way to undo this; it
lasts until the end of the current $\tlang$ component.  
If $\tal{q}$ is $\tal{i}$, $\tal{\mathtt{protect}}$ should not be allowed to hide the
$\tal{i}$th stack slot in $\tal\zeta$; this is enforced by the
restrictions on $\tal{q}$ (see \S\ref{sec:tlang}) when typing $\tal{I}$.

The other new instruction is the $\tlang$ boundary instruction
$\tal{\mathtt{import}}$. Ignoring stacks, the rule is quite simple: it
takes an $\flang$ term $\src{e}$ of type $\src\tau$, well typed under
the $\externmark$ return marker, and translates it to type
$\tytrans\SRC\TAL\tau$, storing the result in register
$\tal{r_d}$. This story is complicated by the handling of stacks, as
it is important for $\tal{\mathtt{import}}$ instructions to be able to
restrict what portion of the stack the inner code can modify. In 
particular, since the $\flang$ code does not have the same return marker
$\tal{q}$, we must be sure that $\tal{q}$ cannot be clobbered by
$\tlang$ code embedded in $\src{e}$. To do this, we specify the
portion of the stack $\tal{\sigma_0}$ that is abstracted as
$\tal\zeta$ in $\src{e}$, and ensure that either $\tal{q}$ is stored
in that stack tail or it is the halting marker. Finally, since the front
of the stack could grow or shrink to $\tal{k}$ entries, if $\tal{q}$
were a stack index $\tal{i}$ we increment it by $\tal{k-j}$ using
the metafunction $\mathrm{inc}$, which otherwise is identity.

\begin{figure}[t]
  \begin{small}
    \[
      \begin{array}{l}
        
    \transval>\TAL\SRC \int {n \withmem {\tal{M}}} = (\tal{n}, \tal{M}) \\
\transval>\TAL\SRC {\mu\alpha.\tau} {\fold {\mu\alpha.\tau} v \withmem
{\tal{M}}} =
  (\tal{\fold{\tytrans\SRC\TAL{\mu\alpha.\tau}} v}, \tal{M'}) \\
\qquad \text{where }
  \transval>\TAL\SRC {\tau[\mu\alpha.\tau/\alpha]} {v \withmem {\tal{M}}}
  = (\tal{v}, \tal{M'}) \\

        \transval>\TAL\SRC {\tuplet {\tau_1, \ldots, \tau_n}}
        {{\tuple{v_0, \ldots, v_n}} \withmem {\tal{M}}} = \\
        \qquad (\tal \ell, (\tal{M_{n+1}}, \tal{\ell} \mapsto \tal{\tuple{w_0, \ldots, w_n}})) \\
\qquad \text{where } \tal{M_0} = \tal{M}, \text{ and }
\transval>\TAL\SRC {\tau_i} {v_i \withmem {\tal{M_i}}}
= (\tal{w_i}, \tal{M_{i+1}}) \\
            \transval>\TAL\SRC \unitt {\unit \withmem {\tal{M}}} = (\tal{\unit}, \tal{M}) \\
        \transval>\TAL\SRC {\arrow{\vec{\tau}}{\tau'}}
  {\src{\func{\vec{x\:\tau}}{t}} \withmem {\tal{M}}} = (\tal{\ell}, (\tal{M}, \tal{\ell} \mapsto \tal{h})) \\
\text{where } \begin{stackTL} \tal{h} = \begin{stackTL}
\tal{\code { \zeta, \epsilon}
           {\reg a \:
            \stopt {\reg1}
                   {\tytrans\SRC\TAL{\tau'}}
                   \zeta
                   \epsilon}
           {\cons {\vec{\tytrans\SRC\TAL \tau }}
                  \zeta}
           {\reg a} {}} \\
\quad  \tal{\salloc 1; \sst 0 {\reg a};
            \import {r_1} {\zeta}
            {\bound>\TAL\SRC {\tau'} e};}\\
          \quad \tal{\sld {\reg a} 0; \sfree {n{+}1}; \retinstr {\reg a}
                  {{\reg 1}}}
\end{stackTL} \\
\src{e} =
\src{(\func{\vec{x\:\tau}}{t})\vec{\bound<\SRC\TAL {\tau}{}}
        (\begin{stackTL}\src{\vec{\tal{\sld {\reg1} {n{+}1{-}i};}}}\\
                        \src{\vec{\tal{\haltinstr
                          {\tytrans\SRC\TAL
                                      \tau}
                                    \sigma
                          {\reg1}},\cdot)}}\end{stackTL}} \\
\tal{\sigma} =
\tal{\cons {\stopt {\reg1}
                   {\tytrans\SRC\TAL{\tau'} }
                   \zeta
                   \epsilon}
           {\cons {\vec{\tytrans\SRC\TAL \tau }}
                  \zeta}} \end{stackTL} \\ 

        \transval<\SRC\TAL \unitt {\unit \withmem {\tal{M}}} = (\src{\unit},
    \tal{M}) \\
    \transval<\SRC\TAL \int {n \withmem {\tal{M}}} = (\src{n}, \tal{M}) \\
\transval<\SRC\TAL {\mu\alpha.\tau}
                    {\fold {\tytrans\SRC\TAL{\mu\alpha.\tau}} w} =
(\src{\fold {\mu\alpha.\tau} v}, \tal{M'}) \\
\qquad \text{where }
\transval<\SRC\TAL {\tau[\mu\alpha.\tau/\alpha]} {w \withmem {\tal{M}}}
= (\src{v},\tal{M'}) \\

        \transval<\SRC\TAL {\tuplet {\tau_0, \ldots, \tau_n}}
{\ell \withmem {\tal{M}}} =
(\src{\tuple{v_0, \ldots, v_n}}, \tal{M_{n+1}}) \\
\qquad \text{where } \begin{stackTL}
\tal{M}(\tal{\ell}) = \tal{\tuple{w_0, \ldots, w_n}}, \\
\tal{M_0} = \tal{M}, \text{ and }
\transval<\SRC\TAL {\tau_i} {w_i \withmem {\tal{M_i}}} = (\src{v_i},\tal{M_{i+1}}) 
\end{stackTL}\\

\transval<\SRC\TAL {\arrow{\vec{\tau_n}}{\tau'}}
{w \withmem {\tal{M}}} = (\src{v},
(\tal{M},\tal{\ell_{end}} \mapsto \tal{h_{end}})) \\
\text{where } \begin{stackTL}
\src{v} = \src{\func{\vec{x_n\:\tau_n}}
                     {\bound<\SRC\TAL {\tau'}
                       {(\begin{stackTL}\tal{\protec{\cdot}{\zeta};} \\
                       \hspace{-2cm}\tal{\import {\reg1} \zeta
                         {\bound>\TAL\SRC {\tau_1} {x_1}}; \salloc 1; \sst 0 {\reg1}; \ldots}\\
                       \hspace{-2cm}\tal{\import {\reg1} \zeta
                                      {\bound>\TAL\SRC {\tau_n} {x_n}};
                              \salloc 1; \sst 0 {\reg1};} \\
                       \hspace{-2cm}\tal{\mv {\reg a}
                         {\ell_{end}[\zeta]};
                         \call w \zeta
                           {\retmark {\tytrans\SRC\TAL {\tau'}} \zeta}},\cdot)\end{stackTL}}}}
\end{stackTL} \\
\tal{h_{end}} = \begin{stackTL}\tal{\code {\zeta}
                           {\reg1 \: {\tytrans\SRC\TAL {\tau'}}}
                           \zeta
                           {\retmark {\tytrans\SRC\TAL {\tau'}} \zeta}{}}\\
                           \quad \tal{\haltinstr {\tytrans\SRC\TAL {\tau'}} \zeta {\reg1}}\end{stackTL}
        \end{array}
      \]
    \caption{\protect\ftlang~Boundary Value Translation}
    \label{ft-value-translation}
  \end{small}
 \vspace{-0.25cm}
\end{figure}

\paragraph{Operational Semantics}

The operational semantics for boundary terms, shown in
Figure~\ref{ft-boundary-semantics}, translate values using the
type-directed metafunctions $\transval<\SRC\TAL \tau \cdot$ ($\tlang$
inside, $\flang$ outside) and $\transval>\TAL\SRC \tau \cdot$
($\flang$ inside, $\tlang$ outside).

Figure~\ref{ft-type-translation} contains the type translation guiding
these metafunctions. Note that $\flang$ tuples are translated to
immutable references to $\tlang$ heap tuples. The most complex
transformation is for function types, which are translated into code
blocks that pass arguments on the stack and follow the
calling convention described in \S\ref{sec:tlang} where return
continuations can be instantiated alternately by $\tlang$ or
$\flang$ callers.

We show the value translations in Figure~\ref{ft-value-translation},
eliding only the stack-modifying lambda, which is similar to the
lambda shown. The most significant translations are between $\tlang$
code blocks and $\flang$ functions. In particular, we must translate
between variable representations and calling conventions---this
means the arguments are passed on the stack, and a return continuation
must be in register $\tal{\reg a}$. Finally, we must translate the
arguments themselves, and translate the return value back, cleaning up
temporary stack values.

Critically, when translating an $\flang$ function to a $\tlang$ code
block, we must protect the return continuation, since embedded
assembly blocks within the body of the function could write to
register $\tal{\reg a}$. To do that, we store $\tal{\reg a}$'s
contents on the stack and protect the tail. In the stack-modifying
lambda case, this is 
complicated slightly by needing to re-arrange the stack to put the
protected value past the exposed stack prefix $\tal{\pref_i}$.
To evaluate the $\flang$ function, we load each argument from
the stack, translate it to $\flang$, apply the function, and import
the returned value back to $\tlang$. After doing this, we load the
return continuation off of the stack, clear the arguments according to
the calling convention and return. Note that in the stack-modifying
lambda case, we have to be careful to clear the arguments but keep the
output prefix $\tal{\pref_o}$.

\paragraph{Example}\label{sec:ft-examples}
In Figure~\ref{higher-order}, we present an example of the type of
transformation that a JIT compiler could perform and the resulting higher-order
callbacks that appear in the multi-language program. At the top of the figure
is the $\flang$ source program which has three functions: $\src{g}$ passes
$\src{1}$ to its argument, $\src{h}$ doubles its argument, and $\src{f}$ passes
$\src{h}$ to its argument. The functions themselves are intentionally minimal,
but we assume the JIT compiler determined that $\src{f}$ and $\src{h}$ should be
compiled to assembly and present the transformed program in the lower half of
the figure. Here, $\src{f}$ and $\src{h}$ have been replaced by code blocks pointed
to by $\tal{\ell}$ and $\tal{\ell_h}$ respectively. 

We present a control-flow diagram for the transformed program in
Figure~\ref{control-higher-order}, where arrows in $\flang$ boxes correspond to
argument passing and return values, whereas arrows in $\tlang$ boxes 
correspond to jumps or halt (as in 
Figure~\ref{control-call-to-call}). 

In this example, when control passes to $\tal{\ell}$, which was compiled from
$\src{f}$, we need to be able to call back into the high-level code in
$\src{g}$. In the block pointed to by $\tal{\ell}$, 
according to the calling convention, the argument $\src{g}$ is passed
on the top of the stack. This 
means that to call back to it, we load it off the stack into register $\tal{\reg
  1}$ with instruction $\tal{\sld{\reg 1}{0}}$, and then $\tal{\mathtt{call}}$
it, as shown in the control-flow diagram in the transfer from 
box $\tlangbox{\vphantom{\tal\ell \src{g}}\tal{~\ell~}}$ to
box $\flangbox{\vphantom{\tal\ell \src{g}}\raisebox{1pt}{\src{~g~}}}$.

But in this example, and indeed in any JIT for higher-order languages, we may
not only need to call from compiled assembly to the interpreted
language, but also be able to pass compiled code back as arguments to the
interpreted language. In this example, the $\tal{\ell_{h}}$ component, which was
compiled from $\src{h}$, is passed as an argument to $\src{g}$. The
function $\src{g}$ then calls $\tal{\ell_{h}}$ with $\src{1}$, causing
control to transfer back to $\tal{\ell_h}$ as we can see in the
transfer to the innermost block in the control-flow diagram. 

The value translation (shown in Figure~\ref{ft-value-translation}) introduces extra blocks where needed, colored as
$\graybox{\tal{\ell_{hret}}}$ and $\graybox{\tal{\ell_{ret}}}$ in our diagram.
These are needed because $\tlang$ components jump to continuation blocks,
whereas for control to pass back to $\flang$ they must $\tal{\mathtt{halt}}$,
which these shim-blocks achieve.

Even though small, this example demonstrates how mixed-language programs with
higher-order callbacks arise naturally in the context of JIT compilation. In
the next section, we'll see how we can use our logical relation to prove these
types of programs equivalent, a necessary step for any proof of correctness for
a JIT compiler.

\begin{figure}[h]
\vspace{-0.45cm}
  \begin{small}
    \[
      \begin{array}{l}
        \src{g} = \src{\func{h \: \arrow{\int}{\int}}{\app{h}{1}}} \\
        \src{h} = \src{\func{x \: \int}{x * 2}}\\
        \src{f} = \src{\func{g \: \arrow{\arrow{\int}{\int}}{\int}}{\app{g}{h}}} \\
        \src{e} = \src{\app{f}{g}}
      \end{array}\hspace{3.5cm}
    \]
  \end{small}

  \vspace{-0.5cm}

  \noindent\rule[2pt]{3cm}{0.33pt} $\downarrow$ JIT Compile $\downarrow$ \rule[2pt]{3cm}{0.33pt}

  \vspace{-0.5cm}
  
  \begin{small}
    \[
      \begin{array}{l}
        \src{\tau} = \src{\arrow{\arrow{\int}{\int}}{\int}} \\
        \src{g} = \src{\func{h \: \arrow{\int}{\int}}{\app{h}{1}}} \\
        \src{e} = \src{(\bound<\SRC\TAL {\src\int} {(\mv {\reg 1} {\ell};
                \haltinstr {\tytrans\SRC\TAL {\arrow{\tau}{\int}}} {\bullet} {\reg 1},
                \tal{H})})}\, \src{g} \\
        \tal{H(\ell)} =
        \begin{stackTL}
                  \tal{\code { \zeta, \epsilon}
                    {\reg a \:
                      \stopt {\reg1}
                      {\tytrans\SRC\TAL{\int}}
                      \zeta
                      \epsilon}
                    {\cons {\tytrans\SRC\TAL {\tau}}
                      \zeta}
                    {\reg a} {}}\\
                  \quad \tal{\sld {\reg 1} 0; \salloc 1; \mv {\reg 2} {\ell_{h}}; \sst 0 {\reg 2};} \\
                  \quad \tal{\sst 1 {\reg a}; \mv {\reg a}
                    {\ell_{gret}[\zeta,\epsilon]};} \\
                  \quad \tal{\callinstr {\reg 1} {\cons{\stopt {\reg1}
                      {\tytrans\SRC\TAL{\int}}
                      \zeta
                      \epsilon}\zeta} {0}}\\
                \end{stackTL} \\
        \tal{H(\ell_{h})} =
        \begin{stackTL}
                  \tal{\code { \zeta, \epsilon}
                    {\reg a \:
                      \stopt {\reg1}
                      {\tytrans\SRC\TAL{\int}}
                      \zeta
                      \epsilon}
                    {\cons {\tytrans\SRC\TAL {\int}}
                      \zeta}
                    {\reg a} {}}\\
                  \quad   \tal{\sld {\reg 1} 0; \sfree 1; \binop{\mathtt{mul}}{\reg 1}{\reg 1}{2}; \retinstr {\reg a} {\reg 1}}
                \end{stackTL} \\
        \tal{H(\ell_{gret})} = 
         \begin{stackTL}
                  \tal{\code{\zeta, \epsilon}
                    {\reg 1 \: \int}
                    {\cons{\codet{}
                    {\reg 1 \: \tytrans\SRC\TAL\int}
                    \zeta
                    \epsilon}{\zeta}}
                    0 {}}\\
                  \quad \tal{\sld {\reg a} 0; \sfree 1; \retinstr {\reg a} {\reg 1}}
                \end{stackTL} \\
        
      \end{array}
    \]
    \caption{\protect\ftlang~Example: JIT}
    \label{higher-order}
  \end{small}
  \vspace{-0.25cm}
\end{figure}

\begin{figure}[h] 
  \begin{small}

    \begin{tikzpicture}
      \node (e) [fblock, align=left] at (0,0) {
        \begin{tikzpicture}
          \node [text width=0cm, align=right] at (5,5.5) {$\flang$};
          \node (lblock) [ftalblock] at (1.8,0) {
            \begin{tikzpicture}
              \node [text width=0cm, align=right] at (5.5,5) {$\tlang$};
              \node (l) [talblock] at (4.5,5) {$\tal{\ell}$};
              \node (geval) [talfblock] at (2.5,0) {
                \begin{tikzpicture}
                  \node [text width=0cm, align=right] at (3,1) {$\flang$};
                  \node (gdef) [fblock, align=left] at (0,1) {$\src{g}$};
                  \node (lh) [ftalblock, align=left] at (0,-2) {
                    \begin{tikzpicture}
                      \node [text width=0cm, align=right] at (2.5,-2.5) {$\tlang$};
                      \node [blank] at (3,1) {};
                      \node (lhdef) [talblock, align=left] at (-1,0.5) {$\tal{\ell_h}$};
                      \node (lhret) [genblock, align=left] at (-0.5,-2.25) {$\tal{\ell_{hret}}$};
                      \node [blank] at (2.5,-2.5) {};

                      \draw [arrow] (current bounding box.north east) to [out=225, in=0]
                      node[memory, pos=0.55] {\calledge{call}{\reg a \map \ell_{hret}}{\cns{1}{\cns{\ell_{gret}}{\cns{\ell_h}{\cns{\ell_{ret}}{\bullet}}}}}}
                      (lhdef);
                      
                      \draw [arrow] (lhdef) to [out=270, in=180]
                      node[memory, pos=0.4] {\calledge{ret}{\reg 1 \map 2}{\cns{\ell_{gret}}{\cns{\ell_h}{\cns{\ell_{ret}}{\bullet}}}}}
                      (lhret);

                      \draw [arrow] (lhret) to [out=0,in=180]
                      node[memory] {\calledge{halt}{\reg 1 \map 2}{\cns{\ell_{gret}}{\cns{\ell_h}{\cns{\ell_{ret}}{\bullet}}}}}
                      (current bounding box.east);
                    \end{tikzpicture}
                  };

                  \node [blank] at (0,-4) {}; 
                  \node [blank] at (0,1.5) {};
                  \draw [arrow] (current bounding box.north west) to [out=300, in=180]
                  node[memory] {\fcall{[\tal{\ell_h}]}}
                  (gdef);
                  \draw [arrow] (gdef.east) to [out=0, in=45]
                  node[memory] {\fcall{\src{1}}}
                  (lh.north east);
                  \draw [arrow] (lh.east) to [out=0, in=135]
                  node[memory] {\fcall{2}}
                  (current bounding box.south east);
                \end{tikzpicture}
              };
              \node (lgret) [talblock, align=left] at (4,-4.5) {$\tal{\ell_{gret}}$};
              \node (lret) [genblock, align=left] at (1,-3.5) {$\tal{\ell_{ret}}$};
              \node [blank] at (-1,-4.5) {};

              \draw [arrow, anchor=north west] (current bounding box.north west) to [out=300, in=170]
                node[memory, anchor=center, pos=0.6] {\calledge{call}{\reg a \map \ell_{lret}}{\cns{[\src{g}]}{\bullet}}}
              (l);

              \draw [arrow] (l) to [out=270, in=90]
              node[memory, pos=0.3] {\calledge{call}{\reg a \map \ell_{gret}}{\cns{\ell_h}{\cns{\ell_{ret}}{\bullet}}}}
              (geval.north west);

              \draw [arrow] (geval.south east) to [out=270, in=45]
              node[memory] {\calledge{ret}{\reg 1 \map 2}{\cns{\ell_{ret}}{\bullet}}}
              (lgret);

              \draw [arrow] (lgret) to [out=180, in=0]
              node[memory] {\calledge{ret}{\reg 1 \map 2}{\bullet}}
              (lret);

              \draw [arrow] (lret) to [out=180, in=45]
              node[memory] {\calledge{halt}{\reg 1 \map 2}{\bullet}}
              (current bounding box.south west);

            \end{tikzpicture}
          
          };
          \node [blank] at (0,4) {};
          \node [blank] at (-2.5,-5.5) {};
          
          \draw [arrow] (current bounding box.north west) to [out=270, in=135]
            node[memory] {\fcall{g}}
            (lblock.north west);

          \draw [arrow] (lblock.south west) to [out=225, in=45]
            node[memory] {\fcall{2}}
            (current bounding box.south west);
          
          \end{tikzpicture}
        };
        
      \end{tikzpicture}
    
    \caption{\protect\ftlang~Control Flow: JIT (Fig.~\ref{higher-order})}
    \label{control-higher-order}
  \end{small}
 \vspace{-0.35cm}
\end{figure}


\section{Logical Relation for \protect\ftlang}
\label{sec:lr}

\begin{figure*}[ht!]
  \begin{small}

    \begin{tabular}{l|l}
      Statement & Meaning \\
      \hline
      $(W,\src{v_1},\src{v_2}) \in \interp V {\src{\tau}} \rho$ & $\src{v_1}$ and $\src{v_2}$ are related $\flang$ values at type $\src{\tau}$ in world $W$ under type substitution $\rho$ \\
      $(W,\tal{w_1},\tal{w_2}) \in \interp W {\tal\tau} \rho$ & $\tal{w_1}$ and $\tal{w_2}$ are related $\tlang$ word values at type $\tal\tau$ in world $W$ under type substitution $\rho$ \\
      $(W,\tal{h_1},\tal{h_2}) \in \interp {HV} {\tal\psi} \rho$ & $\tal{h_1}$ and $\tal{h_2}$ are related $\tlang$ heap values at type $\tal\psi$ in world $W$ under type substitution $\rho$ \\
      $(W,e_1,e_2) \in \obs$ & $e_1$ and $e_2$ run with memories
                               related at $W$, either both terminate
                               or are both running after $W.k$ steps \\
      $(W,E_1,E_2) \in \Krel {q} {\tau} \sigma \rho$ & $E_1$ and $E_2$ are related continuations, so given appropriately related values at type $\tau$, they are in $\obs$ \\
      $(W,e_1,e_2) \in \Erel q \tau \sigma \rho$ & $e_1$ and $e_2$ are related expressions, so given appropriate related continuations, they are in $\obs$ \\ 
    \end{tabular}
    
    \caption{\protect\ftlang~Logical Relation: Closed Values and Terms}
    \label{logical-relation}
  \end{small}
 \vspace{-0.25cm}
\end{figure*}

In order to reason about program equivalence in $\ftlang$, we design a
step-indexed Kripke logical relation for our language. Our logical
relation builds on that of Dreyer~\etal~\cite{dreyer12} and
Ahmed~\etal~\cite{ahmed09:sdri}, where the Kripke worlds contain \emph{islands}
with state-transition systems that we use to accommodate mutations to
the heap, registers, and stack. From those models, we inherit the
ability to reason about equivalences dependent on hidden mutable
state, though we won't go into detail about that aspect in this
paper.  In this section, we focus on the novel aspects of our logical  
relation, showing how we adapted the earlier models to the setting of
$\ftlang$. In particular, the addition of return markers required
non-trivial extensions to the model.

In our logical relation, for which we show the closed relations in
Figure~\ref{logical-relation}, we have three value relations:
$\interp V {\src\tau}\rho$, $\interp W {\tal\tau}\rho$, and
$\interp {HV} {\tal\psi}\rho$.  These correspond to the three types of
values that exist in $\ftlang$: high-level values, low-level
word-sized values, and low-level heap values, respectively. As usual in these relations,
$\rho$ is a relational substitution for type variables. Further, with
the exception of contexts in the 
$\mathcal{K}$ relation all of our relations are built out of
well-typed terms, though we elide that requirement in these figures.

In a Kripke logical relation, relatedness of values depends on the
state of a world $W$. Some values are related irrespective of
world state; for example, an integer $\tal{n}$ is related to itself in
any world $W$, written $(W, \tal{n}, \tal{n}) \in \interp W
{\tal\int}\rho$. However, the structure of the world
captures key semantic properties about the stack, heap, and registers
in a sequence of \emph{islands} that describe the current state of
memories. Each island expresses invariants on certain parts of memory
by encoding a state-transition system and a memory relation that
establishes which pairs of memories are related in each state. 

Since our logical relation is step-indexed our worlds have an index
$k$, which conveys that the relation captures semantic equivalence of
terms for up to $k$ steps but no information is known beyond 
that. This allows us to avoid circularity when dealing with
recursive types as we can induct on the step index rather than the
structure of the expanding type.

$W' \worldext W$ when $W'$ is a future world of
$W$; to reach it, we may have consumed steps (lowering $k$), allocated additional
memory \mbox{in new islands, or made transitions in islands.}

\begin{figure*}[ht!]
  \begin{small}
    \[
      \begin{stackTL}
      \begin{array}{@{}l@{~}l}
        \Erel q \tau \sigma \rho &=
                                   \{\, \begin{stackTL}
                                     (W, e_1 ,e_2) \mid 
                                     \forall E_1, E_2.\ \begin{stackTL}
                                       (W, E_1, E_2) \in \Krel q \tau \sigma \rho
                                       \implies (W, E_1[e_1], E_2[e_2]) \in \obs \,\}
                                     \end{stackTL} \end{stackTL} \\

        \Krel {out} {\src{\tau}} \sigma \rho &=
                                               \{\, \begin{stackTL}
                                                 (W, E_1, E_2)
                                                 \mid
                                                 \forall W',\src{v_1},\src{v_2}. 
                                                 W' \worldextpub W ~\land~
                                                 (W', \src{v_1}, \src{v_2}) \in \interp V {\src{\tau}} \rho ~\land~
                                                 \currS(W') \Subset \interp S {\tal\sigma} \rho \\
                                                 \implies (W', E_1[\src{v_1}], E_2[\src{v_1}]) \in \obs \,\}
                                               \end{stackTL} \\
        \Krel {\retmark \tau \sigma} {\tal{\tau}} \sigma \rho &=
                                                                \{\, \begin{stackTL}
                                                                  (W, E_1, E_2)
                                                                  \mid
                                                                  \forall W',\tal{r_1},\tal{r_2}.
                                                                  W' \worldextpub W ~\land  \\
                                                                  (\later W', W'.\tal{R_1}(\tal{r_1}), W'.\tal{R_2}(\tal{r_2})) \in
                                                                  \interp W {\tal{\tau}} \rho ~\land~ 
                                                                  \currS(W') \Subset \interp S {\tal\sigma} \rho \\
                                                                  \implies (W', \tal{\inlang0{E_1}[(\haltinstr {\rho_1(\tau)} {\rho_1(\sigma)} {r_1}, \cdot)]},
                                                                  \tal{\inlang0{E_2}[(\haltinstr {\rho_2(\tau)} {\rho_2(\sigma)} {r_2},\cdot)]}) \in \obs \,\}
                                                                \end{stackTL} \\

        \Krel q {\tal{\tau}} \sigma \rho &=
                                           \{\, \begin{stackTL}
                                             (W, E_1, E_2)
                                             \mid (\tal{q} = \tal{r} ~\lor~ \tal{q} = \tal{i}) ~\land~
                                             \forall W',\tal{q'},\tal{r_1},\tal{r_2}.
                                             W' \worldextpub W ~\land \\ 
                                             (\exists \tal{r}. \tal{q'} = \tal r ~\land~
                                             \begin{stackTL}
                                               \retaddr_1(W, \rho_1(\tal q)) = W'.\tal{R_1}(\tal r) ~\land~
                                               \retaddr_2(W, \rho_2(\tal q)) = W'.\tal{R_2}(\tal r) ~\land~\\
                                               \retreg_1(W', \tal r) = \tal{r_1} ~\land~ 
                                               \retreg_2(W', \tal r) = \tal{r_2}) ~\land
                                             \end{stackTL} \\
                                             (\later W', W'.\tal{R_1}(\tal{r_1}), W'.\tal{R_2}(\tal{r_2})) \in
                                             \interp W {\tal{\tau}} \rho ~\land~ 
                                             \currS(W') \Subset \interp S {\tal\sigma} \rho  \\
                                             \implies (W', \tal{\inlang0{E_1}[(\retinstr {\inlang0{\rho_1(\tal{q'})}} {r_1}, \cdot)]},
                                             \tal{\inlang0{E_2}[(\retinstr {\inlang0{\rho_2(\tal{q'})}} {r_2},\cdot)]}) \in \obs \,\}
                                           \end{stackTL} \\
      \end{array}\\
      \begin{array}{c@{~~}c@{\quad}c@{~~}c}
        \retaddr_j(W, \tal r) = W.\tal{R_j}(\tal r) & \retaddr_j(W, \tal i) = W.\tal{S_j}(\tal i) &
        \retreg_j(W, \tal r) = \tal{r'} & \mbox{if~} W.\tal{\chi_j}(\tal r) = \tal{\boxof{\stopt {r'} \tau {\sigma'} q}}
      \end{array}
      \end{stackTL}
    \]
      \[
      \tfull{\tal{\Psi}}{\tal{\Delta};\src{\Gamma}}{\chi}{\sigma} q
         {e_1 \lraprx e_2}{\tau}{\sigma'} ~\defeq~
\begin{stackTL}
\forall W,\gamma, \rho.\ \begin{stackTL}
  W \in \interp H {\tal{\Psi}} ~\land~
  \rho \in \interp D {\tal{\Delta}} ~\land 
  (W,\gamma) \in \interp G {\src{\Gamma}}\rho ~\land
\currR(W) \Subset \interp R {\tal\chi} \rho ~\land~ \\
\currS(W) \Subset \interp S {\tal\sigma} \rho
\implies
(W,\rho_1(\gamma_1(e_1)), \rho_2(\gamma_2((e_2))) \in \Erel q \tau {\sigma'} \rho
\end{stackTL} \end{stackTL}
\]
    \caption{\protect\ftlang~Logical Relation: Component and Continuation
      Relations and Equivalence of Open Terms}
    \label{e-and-k-relation}
  \end{small}
 \vspace{-0.15cm}
\end{figure*}

\begin{figure}[h]
  \vspace{-0.5cm}
  \begin{small}
    \[
      \begin{stackTL}
      \interp {HV} {\tal{\codet \Delta \chi \sigma q}} \rho = \\
~~ \{ 
 (W, \begin{stackTL} \tal{\code{\Delta}
               {\inlang0{\rho_1(\tal{\chi})}}
               {\inlang0{\rho_1(\tal{\sigma})}}
               {\inlang0{\rho_1(\tal{q})}}
               {I_1}}, \\
     \tal{\code{\Delta}
               {\inlang0{\rho_2(\tal{\chi})}}
               {\inlang0{\rho_2(\tal{\sigma})}}
               {\inlang0{\rho_2(\tal{q})}}
               {I_2}})
 \mid \\
 \forall W' \worldext W.\
  \forall \rho^* \in \interp D {\tal\Delta}.\
    \forall \tal\tau, \tal{\sigma'}.\ \\
   \quad \text{ let } \rho' = \rho \cup \rho^* \text{ in } 
   \tal\tau; \tal{\sigma'} =_{\rho'} \typeof(\tal q, \tal\chi, \tal\sigma) ~\land~\\
\quad
  \currR(W') \Subset \interp R {\tal{\chi}} \rho'
  ~\land~
  \currS(W') \Subset \interp S {\tal\sigma} \rho' \\
\quad
  \implies
  (W', \tal{(\inlang0{\rho_1^*(\tal{I_1})},\cdot)},
       \tal{(\inlang0{\rho_2^*(\tal{I_2})},\cdot)}) 
       \in \Erel q {\tal\tau} {\sigma'} {\rho'} \,\} \end{stackTL}
   \end{stackTL}
 \]
\[
  \begin{stackTL}
\tal\tau; \tal{\sigma'} =_\rho \typeof(\tal q, \tal\chi, \tal\sigma)
~\defeq~ \\
\quad 
\rho_i(\tal\tau); \rho_i(\tal{\sigma'}) = 
\typeof(\rho_i(\tal q), \rho_i(\tal\chi), \rho_i(\tal\sigma)), \text{ for } i \in 1,2
\end{stackTL}
    \]
    \caption{\protect\ftlang~Logical Relation: Code Block}
    \label{codepointer-relation}
  \end{small}
\vspace{-0.25cm}
\end{figure}

A novel aspect of our logical relation is how it formalizes
equivalence of code blocks at code-pointer type
(Figure~\ref{codepointer-relation}).  Our code-pointer logical relation 
is like a function logical relation in that, given
related inputs, it should produce related outputs. Inputs, in this
case, are registers and the stack for which, in a
future world $W'$ with closing type substitution $\rho^*$, we require that
$\currR(W') \Subset \interp R {\tal\chi}\rho'$ and
$\currS(W') \Subset \interp S {\tal\sigma}\rho'$. This means that the
current register files and stacks in world $W'$ are related at
register file typing $\tal\chi$ and stack typing $\tal\sigma$
respectively. Related register files map registers to related values
and related stacks are made up of related values. Stacks are
related at the stack type $\tal\zeta$ if they are related by relational
substitution $\rho'$.

Once we have related inputs, the logical relation should specify that
applying the arguments produces related output expressions. Since the
arguments are present in the registers and on the stack, we simply
state that the instruction sequences $\tal{I_1}$ and $\tal{I_2}$, with
empty heap fragments, are related components in the $\mathcal{E}$
relation under those conditions.  In this, we rely critically on the
return marker $\tal{q}$ to determine the return type $\tal\tau$ and
resulting stack $\tal{\sigma'}$.

The logical relation $\mathcal{E}$ for components has three formal parameters:
$\tal{q}$, $\tal\tau$, and $\tal{\sigma}$. The return marker $\tal{q}$
says where the expression is returning to as described in
\S\ref{sec:tlang}. The return type $\tal\tau$ is the type of
value that is passed to the return continuation in $\tal{q}$, which is
necessary in order to reason about equivalences, because if expressions
don't even produce the same type of value they can't possibly be
equivalent. This type comes from the $\typeof$ metafunction whose
definition is in Figure~\ref{tal-typing}. The output stack type
$\tal{\sigma}$ is also, in a sense, part of the return value and it
is similarly derived from the return marker by the metafunctions.

The component relation
$\interp E {\tal{q} \vdash \tau; \tal{\sigma}} \rho$ and relation for
evaluation contexts
$\interp K {\tal{q} \vdash \tau; \tal{\sigma}} \rho$ are tightly
connected, as is standard for logical relations based on
biorthogonality. In typical biorthogonal presentations,
the definitions would be:
\[
\vspace{-0.25cm}
\small{  
\begin{stackTL}
  \interp K {\tau} = \{(W,E_1,E_2) \mid \begin{stackTL} \forall W'. W' \worldext W \land (W',v_1,v_2) \in \interp V {\tau} \\ \implies (W',E_1[v_1],E_2[v_2]) \in \obs \}\end{stackTL} \\
  \interp E {\tau} = \{(W,e_1,e_2) \mid \begin{stackTL}\forall E_1 E_2. (W,E_1,E_2) \in \interp K {\tau} \\ \implies (W, E_1[e_1], E_2[e_2]) \in \obs \}\end{stackTL}
  \end{stackTL}
}
\]

\noindent The above states that continuations $E_1$ and $E_2$ accepting type $\tau$
related at world $W$ must be such that, given any future world $W'$
and $\tau$ values, plugging in the values results in related
observations. In turn, expressions $e_1$ and $e_2$ of type $\tau$
related at world $W$ must be such that, given related continuations
$E_1$ and $E_2$, $E_1[e_1]$ and $E_2[e_2]$ are observationally
equivalent. Note how the reduction of $e_1$ and $e_2$ to values is
central, since the definition of $E_1$ and $E_2$ tells you only that given
related values they produce related observations. This reduction is normally
captured in ``monadic bind'' lemmas.

Our definitions, in Figure~\ref{e-and-k-relation}, are more involved
but follow a similar pattern. Our relation $\mathcal{E}$ only differs
from the standard one in that the type of a component involves a
return marker $\tal{q}$ and output stack type $\tal\sigma$. 

The continuation relation $\mathcal{K}$ has three cases for different
return marker $\tal{q}$. The case for $\tal\externmark$, which
corresponds to our functional terms, is nearly identical to the
idealized case shown above.  It differs only in requiring 
$\currS(W') \Subset \interp S {\tal\sigma}\rho$, which means that at
the point we plug in the values $\src{v_1}$ and $\src{v_2}$
the stacks must be related at type $\tal\sigma$.

The $\mathcal{K}$ relation for $\tal{\retmark \tau \sigma}$ is
similar, but since this is $\tlang$ code, return values are stored in
registers $\tal{\mathtt{r_i}}$ and the ``value'' being plugged in
is the $\tal{\mathtt{halt}}$ instruction.

The third case, when the return marker is a register $\tal{\mathtt{r}}$ or
a stack position $\tal{i}$, is more involved, though the overall
meaning is still the same as the other cases: in the future,
we will have a value to pass and will plug it into the hole to get related
observations. First, we note that though at points during computation
the return marker can be a stack index $\tal{i}$, when we actually return to the
continuation the return marker must be stored in a register
$\tal{q'}$. We require, however, that the code block being pointed to
by $\tal{q}$ be the same as that pointed to by $\tal{q'}$. Next, we
find the registers $\tal{r_1}$, $\tal{r_2}$ where the return values
will be passed, and ensure that these contain related values. Finally,
we check that the stacks are related at the right type with
$\currS(W') \Subset \interp S {\tal\sigma}\rho$, before saying that
plugging in the returns must yield related observations.

Having described how closed terms are related, we lift this to open
terms with $\lraprx$, shown at the bottom of
Figure~\ref{e-and-k-relation}. We choose appropriate closing type and
term substitutions, where $\interp G {\src\Gamma}\rho$ is a relational
substitution mapping $\flang$ variables to related $\flang$ values,
and then state the equivalence after closing with these substitutions.

We have proven that the logical relation is sound and complete with
respect to $\ftlang$ contextual equivalence (see technical appendix
\cite{patterson17:funtal-tr}). 

\vspace{0.05cm}
\begin{theorem}[Fundamental Property]
If $\tfull{\tal{\Psi}}{\tal{\Delta};\src{\Gamma}}{\chi}{\sigma} q
         {e}{\tau}{\sigma'}$ then \\
$\tfull{\tal{\Psi}}{\tal{\Delta};\src{\Gamma}}{\chi}{\sigma} q
         {e \lraprx e}{\tau}{\sigma'}$.
\end{theorem}
\vspace{0.05cm}

As usual, we prove compatibility lemmas corresponding to typing rules,
after which the fundamental property follows as a corollary. While 
none of the compatibility lemmas for $\tlang$ instructions are
trivial, the one for $\tal{\mathtt{call}}$ is the most involved.
In particular, $\tal{\mathtt{call}}$ must ensure that the
code that it is jumping to eventually returns, even while the target
component could make nested calls. This relies on the target
component return marker ensuring that control will eventually pass to
the original return continuation.

\vspace{0.05cm}
\begin{theorem}[LR Sound \& Complete wrt Ctx Equiv]
$\tfull{\tal{\Psi}}{\tal\Delta;\src{\Gamma}}{\chi}{\sigma} q
{e_1 \lraprx e_2}{\tau}{\sigma'}$ if and only if \\
$\tfull{\tal{\Psi}}{\tal\Delta;\src{\Gamma}}{\chi}{\sigma} q
{e_1 \ctxeqv e_2}{\tau}{\sigma'}$.
\end{theorem}

\subsection{Example Equivalences}
\label{sec:lr-examples}

In Figure~\ref{basic-blocks}, we show two programs that differ in the
number of basic blocks that they use to carry out the same
computation: adding two to a number and returning it. This example
demonstrates our ability to reason over differences in internal jumps,
which critically depends on the return markers explained in
\S\ref{sec:tlang}. We are able to show these two examples
equivalent at type $\src{\arrow{\int}{\int}}$ using the logical
relation. The elided proofs are included in the technical appendix
\cite{patterson17:funtal-tr}.  

\begin{figure}
  \begin{small}
    \[
\hspace*{-0.25cm}
\begin{array}{@{~}l@{~}l}
  \src{f_1} & = \src{\func{x : \int}{}} \src{\app{\bound< \SRC \TAL
        {\arrow{\int}{\int}}{}
        \hspace{-3pt}(\begin{stackTL}\tal{\protec \cdot \zeta; \mv {\reg 1} {\ell};}\\
          \tal{\haltinstr {\tytrans\SRC\TAL{\src{\arrow{\int}{\int}}}} {\tal\zeta} {\reg 1},}\\
          \tal{H_1}) \hspace{3pt} \src{x} \end{stackTL}
        } {}}\\
   \tal{H_1}(\tal{\ell}) & =
 \begin{stackTL}
 \tal{\code { \zeta, \epsilon}
           {\reg a \:
            \stopt {\reg1}
                   {\tytrans\SRC\TAL{\int}}
                   \zeta
                   \epsilon}
           {\cons {\tytrans\SRC\TAL \int }
                  \zeta}
                {\reg a} {}}\\
  \quad   \begin{stackTL}\tal{\sld {\reg1} 0;
               \binop{\mathtt{add}}{\reg1}{\reg1}{1};
               \binop{\mathtt{add}}{\reg1}{\reg1}{1};} \\
               \tal{\sfree 1;
               \retinstr {\reg a}{{\reg 1}}}\end{stackTL}
\end{stackTL}\\
  \src{f_2} & = \src{\func{x : \int}{}}\src{{\app{\bound< \SRC \TAL {\arrow{\int}{\int}}{}
        \begin{stackTL}(\begin{stackTL}\tal{\protec \cdot \zeta; \mv {\reg 1} {\ell};} \\
          \tal{\haltinstr {\tytrans\SRC\TAL{\src\int}} {\tal\zeta} {\reg 1}},
          \tal{H_2}) \hspace{3pt} \src{x}\end{stackTL}
        \end{stackTL}} {}}}\\
   \tal{H_2}(\tal{\ell}) & =
 \begin{stackTL}
 \tal{\code { \zeta, \epsilon}
           {\reg a \:
            \stopt {\reg1}
                   {\tytrans\SRC\TAL{\int}}
                   \zeta
                   \epsilon}
           {\cons {\tytrans\SRC\TAL \int }
                  \zeta}
                {\reg a} {}}\\
  \quad   \tal{\sld {\reg1} 0;
               \binop{\mathtt{add}}{\reg1}{\reg1}{1};
               \sst 0 {\reg1};
               \jmp{\ell'}[\zeta][\epsilon]}
\end{stackTL}
\\
 \tal{H_2}(\tal{\ell'}) & =
 \begin{stackTL}
 \tal{\code { \zeta, \epsilon}
           {\reg a \:
            \stopt {\reg1}
                   {\tytrans\SRC\TAL{\int}}
                   \zeta
                   \epsilon}
           {\cons {\tytrans\SRC\TAL \int }
                  \zeta}
                {\reg a} {}}\\
  \quad   \tal{\sld {\reg1} 0;
    \binop{\mathtt{add}}{\reg1}{\reg1}{1};
    \sfree 1;
    \retinstr {\reg a}{{\reg 1}}}
\end{stackTL}  
\end{array}
\]
\vspace{-0.25cm}
    \caption{\protect\ftlang~Example: Different Number of Basic Blocks}
    \label{basic-blocks}
  \end{small}
 \vspace{-0.25cm}
\end{figure}

\begin{figure}
  \vspace{-0.25cm}
  \begin{small}

    \[
      \begin{array}{ll}
  \src{fact_F} & = \src{\func{x:\int}
  {\app{(\app{F}{(\fold{\mu
\alpha.\arrow{\alpha}{\int}}{F})})}{x}}}\\
   
      \src{F} & = 
     \begin{stackTL} 
      \src{\func{f:\mu
      \alpha.\arrow{\alpha}{\int}}
      {\func{x:\int}{}}} \\
      \quad \src{\test{x}{1}{(\app{(\app{(\unfold{f})}{f})}{(x-1)})*x}}\\
    \end{stackTL} \\
  \src{fact_T} & = \begin{stackTL}\src{\func{x : \int}
                 {\app{\bound< \SRC \TAL
                 {\arrow{\int}{\int}}{}}}{}}(\\
                 \begin{stackTL}\qquad\tal{\protec \cdot \zeta; \mv {\reg 1} {\ell};} \\
          \qquad\tal{\haltinstr {\tytrans\SRC\TAL{\src\int}} {\tal\zeta} {\reg 1}},\\
          \qquad\tal{H_2}) \hspace{3pt} \src{x}\end{stackTL}
        \end{stackTL}\\ 
   
     \tal{H}(\tal{\ell_{fact}}) & = 
 \begin{stackTL}
 \tal{\codesplit { \zeta, \epsilon}
           {\reg a \:
            \stopt {\reg1}
                   {\tytrans\SRC\TAL{\int}}
                   \zeta
                   \epsilon}
           {\cons {\tytrans\SRC\TAL \int }
                  \zeta}
                {\reg a} {}}\\
  \quad   \begin{stackTL}\tal{\sld {\reg 3} 0;
    \mv {\reg 7} 1;
    \bnz{\reg 3}{\ell_{loop}[\zeta][\epsilon]};} \\
    \tal{\sfree 1;
      \retinstr {\reg a} {\reg 7}}
    \end{stackTL}
\end{stackTL}  
\\
 \tal{H}(\tal{\ell_{loop}}) & = 
 \begin{stackTL}
 \tal{\codesplit { \zeta, \epsilon}
           {\reg 3 \: \tal\int,
             \reg 7 \: \tal\int,\\
           \reg a \:
            \stopt {\reg1}
                   {\tytrans\SRC\TAL{\int}}
                   \zeta
                   \epsilon
                 }
           {\cons {\tytrans\SRC\TAL \int }
                  \zeta}
                {\reg a} {}}\\
  \quad   \begin{stackTL}\tal{\binop{\mathtt{mul}}{\reg 7}{\reg 7}{\reg 3};
           \binop{\mathtt{sub}}{\reg 3}{\reg 3}{1};}\\
           \tal{\bnz{\reg 3}{\ell_{loop}[\zeta][\epsilon]};
           \sfree 1;
             \retinstr {\reg a} {{\reg 7}}}
           \end{stackTL}

\end{stackTL}  
\end{array}
\]
   \vspace{-0.25cm}

    \caption{\protect\ftlang~Example: Factorial Two Different Ways}
    \label{factorial}
  \end{small}
 \vspace{-0.375cm}
\end{figure}

In Figure~\ref{factorial}, we show another small example. We present
two implementations of the factorial function. The $\src{fact_F}$ is a standard
recursive functional implementation using iso-recursive
types. We apply the function template $\src{F}$ to a folded version of
itself and the argument $\src{x}$.  In the body, we check if the
$\src{x}$ is $\src{0}$, in which case we return $\src{1}$, and otherwise we unfold the
first argument, call in with $\src{x-1}$, and multiply the result by
$\src{x}$. This clearly produces the result for $\src{x \ge 0}$, and
also clearly diverges for negative arguments.

The imperative factorial $\src{fact_T}$ uses registers to compute the
result. It has two basic blocks, 
$\tal{\ell_{fact}}$ and $\tal{\ell_{loop}}$. The first, which is
translated to $\flang$ and called with argument $\src{x}$, loads the
argument $\tal{n}$ (translated from $\src{n}$) into register
$\tal{\reg 3}$, stores $\tal{1}$ in the result register
$\tal{\reg 7}$, and then checks if $\tal{\reg 3}$ is $\tal{0}$. If so,
we clear the argument off the stack and return. Otherwise,
we jump to $\tal{\ell_{loop}}$. This multiplies the result by
$\tal{\reg 3}$, subtracts one from $\tal{\reg 3}$, and makes the same
check if $\tal{\reg 3}$ is zero. If so, we do the same cleanup and
return, and otherwise we jump to the beginning of $\tal{\ell_{loop}}$
again.

While these two programs produce the same result, they do it in very
different ways. First, $\src{fact_F}$ uses recursive types, whereas
$\src{fact_T}$ does not. More importantly, $\src{fact_F}$ uses a
functional stack-based evaluation, whereas $\src{fact_T}$ mutates
registers and performs direct jumps. However, the proof of equivalence
only differs from the proof for the example in Figure~\ref{basic-blocks} in that we have to consider two cases --- one in which they
both diverge (for negative input $n$), and one in which they both
terminate with related values (for non-negative input $n$).


\section{Discussion and Future Work}
\label{sec:discussion}

\paragraph{FunTAL for Developers} We have presented a multi-lang\-uage
$\ftlang$ that \emph{safely} embeds assembly in a functional
language. Moreover, our logical relation can be used to establish
correctness of embedded assembly components. Developers of
high-assurance software can write a high-level component
$\src{e}$ to serve as a specification for the TAL implementation
$\tal{e}$ and use our logical relation to prove them equivalent.

$\ftlang$ also enables powerful compositional reasoning about
high-level components, even in the presence of embedded assembly
code. In fact, we conjecture that if the programmer does not use
stack-modifying lambdas and if the embedded TAL contains no
\emph{statically defined} mutable tuples, then $\ftlang$ ensures
referential transparency for high-level terms.  Intuitively, in the
absence of these side-channels (stack-manipulation and mutable cells),
there is no way for two embedded TAL components to communicate with
one another. Thus, even if a high-level term $\src{e}$ contains
embedded assembly, evaluating $\src{e}$ has no observable effects. If
the programmer does use stack-modifying lambdas or statically defined
mutable tuples, reasoning about high-level components remains similar
to reasoning about components in ML.

\paragraph{JIT Formalization}
We plan to investigate modeling a JIT compiler using multi-language
programs. The high-level source language would be untyped and the
low-level language would be typed assembly (since type information is
precisely what a JIT runtime discovers about portions of high-level
code, triggering compilation). We would consider the space of JIT
optimization to be the set of possible replacements of untyped
components with sound low-level versions, with appropriate guards
included to handle violation of typing assumptions. Note, of course,
that the low-level versions may still have calls back into high-level
untyped code. What the JIT is then doing at runtime is moving between
those configurations, usually by learning enough type information to
make the guards likely to pass. 

We can prove a JIT compiler correct based on the transformations that
it would do. Formally, for all moves between configurations that the
JIT may perform, we must show:

\begin{small}
\vspace{-2ex}
\[
  \forall E,\src{e_S}.~ \src{e_S} \overset{E}{\rightsquigarrow} \tal{e_T}
  \implies E\src{[e_S]} \lraprx E\src{[\bound<\SRC\TAL{}{e_T}]} 
\]
\vspace{-3ex}
\end{small}

\noindent where $\overset{E}{\rightsquigarrow}$ represents context-aware
JIT-compilation that allows the compiler to use information in the
context $E$, which could include values in scope, in order to
decide how to transform a component $\src{e_S}$ into $\tal{e_T}$. The
definition of the JIT is thus $\overset{E}{\rightsquigarrow}$, and we
would prove equivalence of the resulting multi-language programs using
a logical relation similar to the one shown in this paper.

\paragraph{Compositional Compiler Correctness} As mentioned in
\S\ref{sec:intro}, Perconti and Ahmed~\cite{perconti14:fca} proved 
correctness of a functional-language compiler that performs closure
conversion and heap allocation.  We can easily adapt our
multi-language to verify correctness of a code-generation pass from their
allocation target $\alang$ to $\tlang$, changing $\ftlang$ to
$\atlang$. The semantics of $\tlang$ and $\tlang$-relevant proofs in
the logical relation can be reused without change.  Correctness of
code generation would then be expressed as contextual equivalence
($\ctxeqv$) in $\atlang$: if $\src{e_A} : \src{\tau_A}$ compiles to
$\tal{e_T}$ then $\src{e_A} \ctxeqv \bound< \SRC \TAL {\tau_A}
{e_T}$.

\paragraph{Continuation-Passing \protect\flang~and Rust} Instead
of trying to bridge the gap between the direct-style $\flang$ and the
continuation-aware $\tlang$, we could have made $\flang$ a
continuation-passing-style language, effectively lowering its level of
abstraction to simplify interoperability with assembly.  But the
resulting multi-language would be more difficult for source
programmers to use, as it would require them to reason about CPS'd
programs. This is essentially the approach taken by the RustBelt
project~\cite{rustbelt16}---i.e., working with a Rust in
continuation-passing style with
embedded unsafe C.\footnote{Personal communication with Derek Dreyer
  and Ralf Jung.}  The project seeks to establish soundness of Rust
and its standard library, where the latter essentially contains unsafe
embedded C. In contrast to $\tlang$, RustBelt does not take a
multi-language approach or aim to handle inline assembly. Rather, it
uses a sophisticated program logic for mutable state to reason about
unsafe C code. It would be interesting to investigate a multi-language
system with direct-style Rust interoperating with unsafe C and
assembly along the lines of our work.

\paragraph{Choices in Multi-Language Design} There are many potential
choices when designing a multi-language system.  For instance, we
chose to expose low-level abstractions to high-level code by adding
stack-modifying lambdas to $\ftlang$ enabling more interactions
between $\flang$ and $\tlang$ code by invalidating equivalences that
might otherwise have been used to justify correctness of compiler
optimizations. We could also add foreign pointers to $\ftlang$, which
would allow references to mutable $\tlang$ tuples to flow into
$\flang$ as opaque values of lump type (as in
Matthews-Findler~\cite{matthews07}), allowing them to be passed but
only used in $\tlang$.  Foreign pointers would have the form
$\bound<\SRC\TAL {L\tuple{\vec{\tau}}} {\tal{\ell}}$ (where
$\tal{\ell} :
\tal{\refto{\tytrans\SRC\TAL{\tuple{\vec\tau}}}}$). While we can
currently provide limited mutation to $\flang$ via $\tlang$ libraries,
foreign pointers would make that more flexible albeit at the cost of
complicating the multi-language.


\section{Related Work}
\label{sec:related}

There has been a great deal of work on multi-language systems, typed 
assembly languages, logics for modular verification of assembly code,
and logical relations in general. We focus our discussion on the most
closely related work. 

Our work builds on results about typed assembly~\cite{morrisett98:popl} and in
particular STAL, its stack-based variant~\cite{morrisett02}.
In \S\ref{sec:tlang} we explain in detail the differences between our TAL and
STAL. Note here though, that these differences stem from our goal to use
type structure to define the notion of a TAL component. We share this goal
with a number of previous type-system design and verification efforts for flavors
of assembly-like languages. 
\citet{glew99:mtal} tackle the problem of safe linking for TAL program
fragments and provide an extension of TAL's type system that
guarantees that linking preserves type safety. 
\citet{benton05:logic-stk} introduces a typed Floyd-Hoare logic
for a stack-based low-level language that treats program fragments
and their linking in a modular fashion. 
Outside the distinct technical details of what a component is in our TAL,
 our work differs from these results in that our
 notion of a TAL component matches that of a function in a high-level
 functional language.

Our multi-language semantics builds of work by Matthews and
Findler~\cite{matthews07} who give multiple interoperability semantics
between a dynamically and statically typed language.  We also build on
multi-languages used for compiler
correctness~\cite{ahmed11:epcps,perconti14:fca,new16:facue} which
embed the source (higher-level) and target (lower-level) languages of
a compiler, though none of that work considers interoperability
with a language as low-level as assembly.

A related strand of research 
explores type safety and foreign function interfaces (FFI).
\citet{furr06} describe sound type inference for the OCaml/C and JNI
FFIs. \citet{tan06:safe-jni} use a mixture of dynamic and static checks to
construct a type-safe variant of JNI.  
\citet{larmuseau15:sec-ffi} aim for fully abstract and type-safe
interoperability between ML and a low-level language. However, their
model low-level language is Scheme with reflection.  
\citet{tan10:jnil} describes a core model for JNI that mixes Java bytecode and
assembly. As an application, they design a sound type system
for their multi-language. Our work is distinct as it captures how
assembly interacts safely with a functional language.


Appel~\etal~showed how to prove soundness of
TALs~\cite{ahmed10:fpcc-jrnl} using (unary) step-indexed models~\cite{ahmed04:phd,ahmed03:impred}. 
Our logical relation most closely resembles the multi-language
relation of Perconti and Ahmed~\cite{perconti14:fca} though theirs, without
assembly, is simpler. Most prior logical relations pertaining to assembly or
SECD machines are cross-language relations that specify equivalence of
high-level (source) code and low-level (target) code and are used to prove
compiler correctness~\cite{benton09,hur11,neis15}. Hur and Dreyer use a
cross-language Kripke logical relation between ML and assembly to verify a
one-pass compiler~\cite{hur11}. Neis~\etal~ set up a parametric inter-language
simulation (PILS) relating a functional source $S$ and a
continuation-passing-style intermediate language $I$, and one relating $I$ to a
target assembly $T$~\cite{neis15}. None of these can reason about equivalence of
(multi-block) assembly components as we do. Jaber and Tabareau~\cite{jaber11}
present a logical relation indexed by source-language types but inhabited by
SECD terms, capturing high-level structure. Besides being able to reason about
mixed programs, our $\ftlang$ logical relation---indexed by multi-language
types---is more expressive: it can be used to prove equivalence of assembly
components of type $\tal{\tau}$ when $\tal{\tau} = \tytrans\SRC\TAL\tau$ for
some $\src{\tau}$ (analogous to Jaber-Tabareau) as well as when $\tal{\tau}$
\emph{is not} of translation type. All of these logical relations make use of
biorthogonality, a natural choice for continuation-based languages.

Finally, Wang~\etal~\cite{wang14} describe a multi-language system in which
components written in a C-like language can link with a simple untyped assembly,
where the latter must be proven to adhere to a specification in higher-order
logic. In their system, equivalences must be specified using axiomatic
higher-order logic specifications. This differs significantly from FunTAL where
equivalences arise out of extensional operational behavior with no external
specification needed. Further, all of their assembly must be proven to follow an
XCAP program specification, making it a much heavier approach than our typed
assembly language. Our approach is complementary, in that while their
higher-order logic allows finer grained specifications, it incurs additional
cost on the programmers, and indeed renders it potentially non-viable for ML and
x86 programmers that we believe would still be able to use FunTAL.



\vspace{-0.9ex}

\acks

This research was supported in part by the National Science Foundation
(grants CCF-1422133, CCF-1453796, CCF-1618732, CCF-1421770, and CNS-1524052) and a Google
Faculty Research Award. 

\balance

\bibliographystyle{abbrvnat}

\bibliography{amal}

\end{document}